\documentclass[12pt,a4paper,notitlepage]{article}

\usepackage{amsfonts}
\usepackage{amsmath}
\usepackage{amssymb}
\usepackage{amsthm}
\usepackage{color}
\usepackage[T1]{fontenc}
\usepackage[utf8]{inputenc}
\usepackage{graphicx}
\usepackage{hyperref}
\usepackage{mathtools}
\hypersetup{
  colorlinks,
  citecolor=blue,
  linkcolor=blue,
  urlcolor=blue}
\usepackage{mathrsfs}
\usepackage{xurl}
\usepackage[square, sort&compress]{natbib}
\usepackage[format=hang]{subcaption}
\usepackage{times}
\usepackage{upgreek}
\usepackage{import}
\usepackage{algorithm2e}
\RestyleAlgo{ruled}
\usepackage{bbm}
\usepackage{mathtools}
\usepackage[nameinlink]{cleveref}
\usepackage{multirow}
\usepackage{multicol}
\usepackage{float}
\usepackage{setspace}
\usepackage{array}
\AtBeginDocument{} 

\usepackage{svg}
\usepackage{pgf}
\usepackage{tikz}
\usepackage{pgfplots}

\pgfplotsset{compat=newest}

\definecolor{rwthblue}{RGB}{0,84,159}      
\definecolor{rwthlightblue}{RGB}{142,186,229}   
\definecolor{rwthpetrol}{RGB}{0,97,101}      
\definecolor{rwth4}{RGB}{0,152,161}     
\definecolor{rwthgreen}{RGB}{87,171,39}     
\definecolor{rwth6}{RGB}{189,205,0}     
\definecolor{rwth7}{RGB}{255,237,0}     
\definecolor{rwthorange}{RGB}{246,168,0}     
\definecolor{rwth9}{RGB}{227,0,102}     
\definecolor{rwthred}{RGB}{204,7,30}     
\definecolor{rwth11}{RGB}{161,16,53}    
\definecolor{rwth12}{RGB}{97,33,88}     
\definecolor{rwth13}{RGB}{122,111,172}  

\date{}

\graphicspath{{./images/}}

\begin{document}
\pgfkeys{/pgf/number format/.cd,1000 sep={}}

\author{\large{Jannick Kehls${}^{\,a}$, Stephan Ritzert${}^{\,a}$, Lars Breuer${}^{\,a}$, Qinghua Zhang${}^{\,a}$,}\\{ Stefanie Reese${}^{\,a}$${}^{\,b}$, Tim Brepols${}^{\,a}$}\\[0.5cm]
\hspace*{-0.1cm}
\normalsize{\em  ${}^{a}$Institute of Applied Mechanics, RWTH Aachen
  University,}\\
\normalsize{\em Mies-van-der-Rohe-Str.\ 1, 52074 Aachen, Germany}\\[0.25cm]
\normalsize{\em ${}^{b}$University of Siegen, 57076 Siegen, Germany}\\
}

\title{\LARGE Multi-field decomposed hyper-reduced order modeling of damage-plasticity simulations}
\maketitle

\small
{\bf Abstract.} This paper presents a multi-field decomposed approach for hyper-reduced order modeling to overcome the limitations of traditional model reduction techniques for gradient-extended damage-plasticity simulations. The discrete empirical interpolation method (DEIM) and the energy-conserving sampling and weighting method (ECSW) are extended to account for the multi-field nature of the problem. Both methods yield stable reduced order simulations, while significantly reducing the computational cost compared to full-order simulations. Two numerical examples are presented to demonstrate the performance and limitations of the proposed approaches. The decomposed ECSW method has overall higher accuracy and lower computational cost than the decomposed DEIM method.

\vspace*{0.3cm}
{\bf Keywords:} {model order reduction, proper orthogonal decomposition, discrete empirical interpolation method, energy-conserving sampling and weighting method, nonlinear solid mechanics}


\normalsize


\section{Introduction}

Modeling damage and plasticity in materials is a challenging task, especially when it comes to simulating large structures with complex geometries. To overcome the well-known pathological mesh dependence in finite element simulations involving damage-induced strain localization, nonlocal and gradient-extended damage models have been developed. Essentially, these models introduce an internal length scale parameter into the formulation that allows for a more distributed representation of damage, helping to avoid the mentioned numerical problems. However, the computational cost of such nonlinear simulations can easily become prohibitively high, particularly when high fidelity models are required to capture the intricate behavior of the considered structure under load.

It is therefore of great interest to develop methods that reduce the computational complexity without loosing the validity of the results. The reduction of problems described by partial differential equations began to increase significantly with the growing use of reduced basis methods, such as the proper orthogonal decomposition (POD) \citep{chatterjee_introduction_2000}. It has since then been applied to various research and engineering problems, such as turbulent flow \citep{berkooz_proper_2003, smith_low-dimensional_2005,hijazi_data-driven_2020}, dynamical systems \citep{benner_proper_2005, kerschen_method_2005,lu_applications_2021} and solid mechanics \citep{chen_proper_2005, ritzert_adaptive_2024,ritzert_componentbased_2025}. A comparative study by Radermacher and Reese \citep{radermacher_comparison_2013} showed that the POD yields very good results when applied to problems in nonlinear solid mechanics and outperforms the modal basis method \citep{spiess_reduction_2005,nickell_nonlinear_1976} as well as the load-dependent Ritz method \cite{idelsohn_reduction_1985, ouanani_sensitivity_2020, koh_ritz_2023}.

Due to the nonlinearity of the partial differential equation at hand, an investigation into hyper-reduction techniques followed, with one of the first contributions being the reconstruction of gappy data \citep{everson_karhunenloeve_1995}. These methods aim to reduce the complexity originating from the repeated computation of the stiffness matrix and the residual vector. In the field of computational mechanics, several methods have been applied with success, such as the a priori hyper-reduction method \citep{ryckelynck_priori_2005-1,ryckelynck_hyper-reduction_2009,miled_priori_2013}, the empirical cubature method \citep{hernandez_dimensional_2017, martinez_local-ecm_2023, hernandez_cecm_2024}, the energy-conserving sampling and weighting (ECSW) method \citep{an_optimizing_2008,farhat_dimensional_2014, farhat_structure-preserving_2015, trainotti_ecsw_2024} and the discrete empirical interpolation method (DEIM) \citep{chaturantabut_nonlinear_2010,radermacher_pod-based_2016,maierhofer_model_2022}.

The DEIM, which will be discussed in this paper, promises higher speedups than the POD alone, but generally at the expense of accuracy. Since its initial development, many alterations to the method have been published. In the finite element setting, an interpolation degree of freedom can lead to the evaluation of many neighboring elements, which is why an unassembled DEIM was proposed by \citet{tiso_modified_2013}, leading to higher reductions in computation time. \citet{peherstorfer_localized_2014} presented the possibility to create localized reduced order models (ROMs) for specific regions of systems. These regions can be clustered by means of machine learning in the offline computation and this alteration typically yields more accurate results. \citet{drmac_new_2016} proposed a new selection procedure for the interpolation points by means of a QR factorization, which has a sharper error bound for the DEIM projection error. Their results showed better accuracy of the ROM, while their approach is relatively easy to implement with freely-available software packages. In this paper, the DEIM is extended to account for multi-field problems. Separate projection matrices are computed for the different fields, which allows for a more accurate representation of the system and counteracts simulation instabilities that can occur using the standard DEIM.

In addition to the extension of the DEIM, the ECSW is extended in this paper to account for multi-field problems. The ECSW was first introduced by \citet{an_optimizing_2008} and has since then been applied to solid mechanics problems by \citet{farhat_dimensional_2014,grimberg_mesh_2021}. The ECSW is a hyper-reduction method that uses a sampling and weighting approach to reduce the computational cost of the problem by approximating the energy of the projection of the internal force vector with a reduced set of elements and corresponding weights. The ECSW has been shown to yield very good results in terms of accuracy and speedup \citep{bodnar_comparison_2023}. However, in combination with the standard POD on which it is based, numerical instabilities can occur in highly nonlinear problems, which might be one of the reasons why it has not yet been applied to gradient-extended damage-plasticity problems. In this paper, the ECSW is extended to account for general multi-field problems. The extension is similar to the one of the DEIM, where separate projection matrices are computed for the different physical fields.

The paper is structured as follows: In \Cref{sec:mat}, the gradient-extended damage-plasticity model is briefly introduced, leading to the final coupled equation system under consideration. In \Cref{sec: MOR}, the decomposed POD, DEIM and ECSW are introduced and explained in detail. In \Cref{sec: numerical examples}, two numerical examples are presented to demonstrate the performance of the methods. The first example focuses on the dependency of the methods on the chosen {(hyper-)}reduction parameters, whereas the second example focuses on the applicability of the methods to an optimization problem. Finally, in \Cref{sec: conclusion}, the results are summarized and an outlook on future work is given.
\section{Material Model}
\label{sec:mat}

In the following, the constitutive model used to investigate the proposed model order reduction approach is briefly introduced. The material model was developed by \citet{brepols_gradient-extended_2020} and serves as the finite strain version of a corresponding small strain model introduced earlier \citep{brepols_gradient-extended_2017}. The material model is based on the micromorphic approach proposed by \citet{forest_micromorphic_2009} which can be seen as a systematic procedure to construct `non-local' gradient-extended material models of already existing `local' counterparts. Essentially, an additional nodal degree of freedom is introduced into the formulation, the so-called micromorphic damage, whose evolution is strongly coupled to a corresponding local internal damage variable via a penalization term in the Helmholtz free energy. This approach is chosen to overcome the general problem of conventional local continuum damage models, which suffer from an artificial dependence of the softening zone on the mesh discretization in a finite element simulation.

As the model is based on finite deformation kinematics, the reference configuration $B_0$ at time $t_0$ and the current configuration $B_t$ at time $t$ of a body are differentiated. In the two configurations, the position of a material point is described by the vectors $\boldsymbol{X}$ and $\boldsymbol{x}$, respectively. The current position of a material point can be obtained by adding the displacement vector $\boldsymbol{u}$ to the position of the material point in the reference configuration, i.e. $\boldsymbol{x} = \boldsymbol{X} + \boldsymbol{u}$. Assuming there exists a mapping $\boldsymbol{x} = \boldsymbol{\chi}(\boldsymbol{X}, t)$, the deformation gradient is introduced as $\mathbf{F} \coloneqq \partial \boldsymbol{\chi}(\boldsymbol{X},t)/\partial \boldsymbol{X}$, the right Cauchy-Green deformation tensor as $\mathbf{C} = \mathbf{F}^T \mathbf{F}$ and the Green-Lagrange strain tensor as $\mathbf{E} = \frac{1}{2}(\mathbf{C} - \mathbf{I})$, where $\mathbf{I}$ is the identity. To model elastoplastic material behavior, the deformation gradient is multiplicatively decomposed into the elastic part $\mathbf{F}_e$ and the plastic part $\mathbf{F}_p$. As the model includes Armstrong-Frederick type non-linear kinematic hardening, the plastic part of the deformation gradient is further decomposed into the recoverable part $\mathbf{F}_{p_e}$ and the irrecoverable part $\mathbf{F}_{p_i}$. Based on these splits, the deformation gradient reads
\begin{equation}
    \mathbf{F} = \mathbf{F}_e \mathbf{F}_p \text{~ with ~} \mathbf{F}_p = \mathbf{F}_{p_e} \mathbf{F}_{p_i}.
\end{equation}
From this, the elastic right Cauchy-Green deformation tensors are derived as
\begin{equation}
    \mathbf{C}_e = \mathbf{F}_e^T \mathbf{F}_e, \qquad \mathbf{C}_{p_e} = \mathbf{F}_{p_e} ^T \mathbf{F}_{p_e}.
\end{equation}
The Helmholtz free energy is assumed to be
\begin{equation}
    \psi = f_{\textrm{dam}}(D)(\psi_e(\mathbf{C}_e)+\psi_p(\mathbf{C}_{p_e}, \xi_p))+\psi_d(\xi_d) + \psi_{\bar{d}}(D-\bar{D}, \mathrm{Grad}(\bar{D}))
\end{equation}
where $\psi_e$ and $\psi_p$ denote the elastic and plastic part of the free energy, respectively. Besides its dependence on the recoverable plastic part of the right Cauchy Green deformation tensor $\mathbf{C}_{p_e}$, the plastic part of the free energy also depends on the accumulated plastic strain $\xi_p$. The energy related to damage hardening is stored in $\psi_d$, whereas the energy related to the micromorphic extension is stored in $\psi_{\bar{d}}$. Furthermore, the damage hardening energy depends on the damage hardening variable $\xi_d$ and the energy related to the micromorphic extension on the micromorphic damage variable $\bar{D}$. Lastly, the scalar-valued weakening function $f_{\textrm{dam}}(D)$ is introduced, which depends on the local damage variable $D$ and needs to be twice-differentiable and monotonically decreasing. A compressible Neo-Hookean type energy is chosen for the elastic part and reads
\begin{equation}
    \psi_e = \dfrac{\mu}{2}\left(\mathrm{tr} \mathbf{C}_e - 3 - 2\ln\left(\sqrt{\mathrm{det}\mathbf{C}_e}\right)\right) + \dfrac{\lambda}{4}\left(\mathrm{det}\mathbf{C}_e -1 -2 \ln\left(\sqrt{\mathrm{det}\mathbf{C}_e}\right)\right)
\end{equation}
with the two Lamé constants $\mu$ and $\lambda$. The plastic energy is defined as
\begin{equation}
    \psi_p = \dfrac{a}{2}\left(\mathrm{tr} \mathbf{C}_{p_e} - 3 - 2\ln\left(\sqrt{\mathrm{det}\mathbf{C}_{p_e}}\right)\right) + e\left(\xi_p + \dfrac{\exp(-f \xi_p)-1}{f}\right).
\end{equation}
The plastic material parameters $a$, $e$ and $f$ are related to kinematic and non-linear Voce isotropic hardening, with another parameter $b$ showing up in the evolution equation of $\dot{\mathbf{C}}_{p_i}$ (see \autoref{eq:Cpidot}).
The energy related to Voce type damage hardening is defined as
\begin{equation}
    \psi_d = r\left(\xi_d + \dfrac{\exp\left(-s\xi_d\right) - 1}{s} \right)
\end{equation}
where $r$ and $s$ are the two damage hardening material parameters. The energy of the micromorphic extension reads
\begin{equation}
    \psi_{\bar{d}} = \dfrac{H}{2}(D-\bar{D})^2 + \dfrac{A}{2} \mathrm{Grad}(\bar{D})\cdot\mathrm{Grad}(\bar{D})
\end{equation}
with a penalty parameter $H$ and the parameter $A$, controlling the influence of the micromorphic damage gradient in the formulation. An internal length scale is introduced into the material as $L \coloneqq \sqrt{A / H}$ which can be understood as a localization limiter, preventing the damage of shrinking into a zone of vanishing volume. Choosing the penalty parameter to be a very high number allows for a strong coupling between the micromorphic and the local damage variable.

Thermodynamically consistent constitutive equations with respect to the reference configuration can be derived based on the Clausius-Duhem inequality and are summarized in the following (details are omitted, but can be found in \citet{brepols_gradient-extended_2020}):

$\bullet$ State laws:
\begin{equation} \begin{aligned}
        \mathbf{S} & = f_{\mathrm{dam}}(D) \, 2 \, \mathbf{F}_p^{-1} \dfrac{\partial \psi_e}{\partial \mathbf{C}_e}\mathbf{F}_p^{-T}                                                                                    \\
                   & = f_{\mathrm{dam}}(D) \left(\mu \, (\mathbf{C}_p^{-1} - \mathbf{C}^{-1}) + \dfrac{\lambda}{2} \left(\dfrac{\mathrm{det} \mathbf{C}}{\mathrm{det} \mathbf{C}_p} - 1 \right) \mathbf{C}^{-1} \right)
    \end{aligned} \end{equation}
\begin{equation}
    a_{0_i} = \dfrac{\partial \psi_{\bar{d}}}{\partial \bar{D}} = -H (D-\bar{D})
    \label{eq:a}
\end{equation}
\begin{equation}
    \boldsymbol{b}_{0_i} = \dfrac{\partial \psi_{\bar{d}}}{\partial \mathrm{Grad}(\bar{D})} = A \, \mathrm{Grad}(\bar{D})
    \label{eq:b}
\end{equation}
$\bullet$ Thermodynamic conjugate forces - plasticity:
\begin{equation}
    \mathbf{X} = f_{\mathrm{dam}}(D) \, 2 \, \mathbf{F}_{p_i}^{-1} \dfrac{\partial \psi_e}{\partial \mathbf{C}_{p_e}}\mathbf{F}_{p_i}^{-T} = f_{\mathrm{dam}}(D)\,a\,(\mathbf{C}_{p_i}^{-1} - \mathbf{C}^{-1})
\end{equation}
\begin{equation}
    q_p = f_{\mathrm{dam}}(D)\,\dfrac{\partial \psi_p}{\partial \xi_p} = f_{\mathrm{dam}}(D)\,e\,(1 - \exp(-f\xi_p))
\end{equation}
$\bullet$ Thermodynamic conjugate forces - damage:
\begin{equation}
    Y = - \dfrac{\mathrm{d} f_{\mathrm{dam}}(D)}{\mathrm{d} D}\,(\psi_e+\psi_p)-\dfrac{\partial \psi_{\bar{d}}}{\partial D} = - \dfrac{\mathrm{d} f_{\mathrm{dam}}(D)}{\mathrm{d} D}\,(\psi_e+\psi_p) - H(D-\bar{D})
\end{equation}
\begin{equation}
    q_d = \dfrac{\partial \psi_d}{\partial \xi_d} = r\,(1 - \exp(-s \, \xi_d))
\end{equation}
$\bullet$ Auxillary stress tensors:
\begin{equation}
    \mathbf{Y} = \mathbf{C} \mathbf{S} - \mathbf{C}_p \mathbf{X}, \quad \mathbf{Y}_{\textrm{kin}} = \mathbf{C}_p \mathbf{X}
\end{equation}
$\bullet$ Evolution equations expressed in terms of effective (i.e., `undamaged') quantities:
\begin{equation}
    \tilde{(\bullet)} \coloneqq \frac{(\bullet)}{f_{\mathrm{dam}}(D)}
\end{equation}
\begin{equation}
    \dot{\mathbf{C}}_p = 2 \, \dot{\lambda}_p \dfrac{\sqrt{3/2}}{f_{\mathrm{dam}}(D)}\dfrac{\tilde{\mathbf{Y}}^{'} \mathbf{C}_p}{\sqrt{\tilde{\mathbf{Y}}^{'} \cdot (\tilde{\mathbf{Y}}^{'})^T}}, \quad \dot{\mathbf{C}}_{p_i} = 2 \, \dot{\lambda}_p \dfrac{b/a}{f_{\mathrm{dam}}(D)} \mathbf{Y}^{'}_{\textrm{kin}} \mathbf{C}_{p_i}
    \label{eq:Cpidot}
\end{equation}
\begin{equation}
    \dot{\xi}_p = \dfrac{\dot{\lambda}_p}{f_{\mathrm{dam}}(D)}, \quad \dot{D} = \dot{\lambda}_d, \quad \dot{\xi}_d = \dot{\lambda}_d
\end{equation}
$\bullet$ Yield and damage loading function:
\begin{equation}
    \Phi_p = \sqrt{3/2} \sqrt{\tilde{\mathbf{Y}}^{'} \cdot (\tilde{\mathbf{Y}}^{'})^T} - (\sigma_0 + \tilde{q}_p), \quad \Phi_d = Y - (Y_0 + q_d)
    \label{eq:threshold}
\end{equation}
$\bullet$ Loading / unloading conditions:
\begin{equation} \begin{aligned}
        \dot{\lambda}_p \ge 0, \qquad \Phi_p \le 0, \qquad \dot{\lambda}_p \Phi_p = 0 \\
        \dot{\lambda}_d \ge 0, \qquad \Phi_d \le 0, \qquad \dot{\lambda}_d \Phi_d = 0
    \end{aligned} \end{equation}
Adding the initial yield stress $\sigma_0$ and damage threshold $Y_0$ in \autoref{eq:threshold}, the material model has twelve material parameters in total ($\lambda$, $\mu$, $\sigma_0$, $a$, $b$, $e$, $f$, $Y_0$, $r$, $s$, $A$, $H$) which should be optimally determined from experimental data.

Based on the generalized stresses $a_{0_i}$ and $b_{0_i}$ in Equations \ref{eq:a} and \ref{eq:b}, respectively, the micromorphic balance equation can be written as
\begin{equation} \begin{aligned}
        H(D-\bar{D})+ A \, \mathrm{Div}(\mathrm{Grad}(\Bar{D})) & = 0 \quad \text{in } B_0          \\
        \mathrm{Grad}(\bar{D}) \cdot \boldsymbol{n}_0           & = 0 \quad \text{on } \partial B_0
    \end{aligned} \end{equation}
Having the micromorphic balance equation at hand as well as the balance of linear momentum
\begin{equation} \begin{aligned}
        \mathrm{Div}(\mathbf{F} \mathbf{S}) + \boldsymbol{f}_0 & = 0 \quad                    &  & \text{in } B_0          \\
        \mathbf{F} \mathbf{S}[\boldsymbol{n}_0]                & = \boldsymbol{t}_0 \quad     &  & \text{on } \partial B_0 \\
        \boldsymbol{u}                                         & = \hat{\boldsymbol{u}} \quad &  & \text{on } \partial B_0
    \end{aligned} \end{equation}
the weak form of the problem can be constructed. This is done via the usual procedure: First, the balance equations are multiplied by the appropriate test functions $\delta \boldsymbol{u}$ in case of the linear momentum and $\delta \bar{D}$ in case of the micromorphic balance and then integrated over the reference domain $B_0$. After that, partial integration and the divergence theorem are applied such that the weak form reads:
\begin{equation}
    g(\boldsymbol{u}, \bar{D}, \delta \boldsymbol{u}) \coloneqq \int_{B_0} \mathbf{S} \cdot \delta \mathbf{E} \: \mathrm{d} V - \int_{B_0} \boldsymbol{f}_0 \cdot \delta \boldsymbol{u} \: \mathrm{d} V - \int_{\partial B_{0_t}} \boldsymbol{t}_0 \cdot \delta \boldsymbol{u} \: \mathrm{d} A = 0 \quad \forall \delta \boldsymbol{u}
\end{equation}
\begin{equation}
    h(\boldsymbol{u}, \bar{D}, \delta \bar{D}) \coloneqq \int_{B_0} (H(D- \bar{D})\delta \bar{D} - A \, \mathrm{Grad}(\bar{D}) \cdot \mathrm{Grad}(\delta \bar{D}))\mathrm{d} V = 0 \quad \forall \delta \bar{D}
\end{equation}
The weak form is generally nonlinear and must be linearized with respect to the unknowns and solved iteratively. Details about the linearization and subsequent finite element discretization can be found in \citet{brepols_gradient-extended_2020}.

Finally, one ends up with a global system of finite element equations which reads
\begin{equation}
    \underbrace{\left[\begin{array}{l l} \mathbf{K}_{UU}  & \mathbf{K}_{U \bar{D}}       \\
                \mathbf{K}_{\bar{D} U} & \mathbf{K}_{\bar{D} \bar{D}}
            \end{array} \right]}_{\mathbf{K}_T}
    \underbrace{\left \{\begin{array}{l} \Delta \boldsymbol{U} \\
            \Delta \bar{\boldsymbol{D}}
        \end{array} \right \}}_{\Delta \boldsymbol{V}}
    = -
    \underbrace{\left \{ \begin{array}{l} \boldsymbol{R}_U - \boldsymbol{P}_U \\
            \boldsymbol{R}_{\bar{D}}
        \end{array} \right \}}_{\boldsymbol{G}}
    \label{eq:soe}
\end{equation}
The abbreviations $\mathbf{K}_T$ for the tangential stiffness matrix, $\Delta \boldsymbol{V}$ for the incremental unknowns vector, and $\boldsymbol{G}$ for the residual vector are introduced for later purposes. To obtain a solution to the problem at hand, the equation system \autoref{eq:soe} is solved iteratively until convergence is achieved.
For further details regarding the individual components of the tangential stiffness matrix $\mathbf{K}_{UU}$, $\mathbf{K}_{U \bar{D}}$, $\mathbf{K}_{\bar{D} U}$, $\mathbf{K}_{\bar{D} \bar{D}}$ and the internal and external force vectors $\boldsymbol{R}_U$, $\boldsymbol{P}_U$, $\boldsymbol{R}_{\bar{D}}$ as well as the algorithmic implementation at the Gauss point level, the reader is again kindly referred to \citet{brepols_gradient-extended_2020}.

Because of the complexity of the material model, the repeated computation of the stiffness matrix and residual vector is fairly expensive. It should be noted at this point that in the present work, full integration with linear interpolation functions and eight integration points per element are used. For completeness, the work of \citet{barfusz_single_2021} should be highlighted, in which a single Gauss point formulation was invented for the gradient-extended material model at hand to prevent spurious locking of the formulation. It was shown that the reduced integration approach still yields accurate results and, as a side effect, naturally increases the computational efficiency. Nevertheless, this gain in computational efficiency by itself is usually not sufficient to achieve the required speedup, for example, for large-scale problems. This issue is tackled in the following thoroughly by means of suitable model order reduction techniques
\section{Decomposed model order reduction}
\label{sec: MOR}

Having specified the nonlinear solid mechanics problem in the previous section, the model order reduction (MOR) techniques that will be used to reduce the computational cost of the problem are explained in the present section. Firstly, the decomposed POD (DPOD) is recapitulated. Following that, the decomposed DEIM (DDEIM) and the decomposed ECSW (DECSW) are introduced and the decomposition with respect to the different fields is explained in detail.

\subsection{Projection-based model order reduction}
\label{sec: pMOR}

A basic assumption of the projection-based MOR is that the solution vector of the system can be approximated by a linear combination of a reduced set of basis vectors, the latter of which are collected in a column-wise manner in the projection matrix $\mathbf{\Phi} \in \mathbb{R}^{n \times m}$, such that
\begin{equation}
    \boldsymbol{V} \approx \mathbf{\Phi} \hat{\boldsymbol{V}},
\end{equation}
where $\hat{\boldsymbol{V}}$ describes the reduced solution vector. Using this approximation, a Galerkin projection of \autoref{eq:soe} leads to the following expression for the reduced linearized and discretized weak form
\begin{equation}
    \underbrace{\mathbf{\Phi}^T \, \mathbf{K}_T(\mathbf{\Phi} \hat{\boldsymbol{V}}) \, \mathbf{\Phi}}_{\coloneqq \hat{\mathbf{K}}_T(\mathbf{\Phi} \hat{\boldsymbol{V}})} \; \Delta \hat{\boldsymbol{V}} = - \underbrace{\mathbf{\Phi}^T \boldsymbol{G}(\mathbf{\Phi} \hat{\boldsymbol{V}})}_{\coloneqq \hat{\boldsymbol{G}}(\mathbf{\Phi} \hat{\boldsymbol{V}})}.
\end{equation}
The reduced iterative solution scheme is described by:
\begin{equation}
    \label{eq:reducedsoe}
    \begin{aligned}
         & \hat{\mathbf{K}}_T(\mathbf{\Phi} \hat{\boldsymbol{V}}^i) \; \Delta \hat{\boldsymbol{V}}^{i+1} = - \hat{\boldsymbol{G}}(\mathbf{\Phi} \hat{\boldsymbol{V}}^i) \\
         & \boldsymbol{V}^{i+1} = \boldsymbol{V}^{i} + \mathbf{\Phi} \Delta \hat{\boldsymbol{V}}^{i+1}                                                                  \\
         & ||\hat{\boldsymbol{G}}(\mathbf{\Phi} \hat{\boldsymbol{V}}^i)|| < tol                                                                                         \\
         & i = i + 1.
    \end{aligned}
\end{equation}
In the above scheme, $i$ denotes the iteration step. The reduced iterative solution scheme reduces the computational cost of the system by reducing the number of degrees of freedom (DOFs) of the system from $n$ to $m$ with the assumption that a projection matrix $\mathbf{\Phi}$ can be obtained such that $m \ll n$ without introducing a too large error.

\subsection{(Decomposed) proper orthogonal decomposition}
\label{sec: DPOD}

One way of constructing the projection matrix $\mathbf{\Phi}$ is to use the POD. The POD basis is constructed by computing the singular value decomposition (SVD) of a snapshot matrix $\mathbf{D} = [\boldsymbol{V}_1 \, \boldsymbol{V}_2 \, \dots \, \boldsymbol{V}_{\ell}] \in \mathbb{R}^{n \times \ell}$, where $n$ is the number of DOFs and $\ell$ is the number of snapshots. The snapshot matrix contains solution vectors that are computed in an offline phase using the full order model. The snapshots can either include only solution states from one simulation or also from multiple simulations with different simulation parameters. The SVD of the snapshot matrix is given by
\begin{equation}
    \mathbf{D} = \mathbf{\Psi} \mathbf{\Sigma} \mathbf{W},
\end{equation}
where $\mathbf{\Psi} = [\boldsymbol{\psi}_1 \, \boldsymbol{\psi}_2 \, \dots \, \boldsymbol{\psi}_n] \in \mathbb{R}^{n \times n}$ and $\mathbf{W} \in \mathbb{R}^{\ell \times \ell}$ are orthogonal matrices and $\mathbf{\Sigma} \in \mathbb{R}^{n \times \ell}$ is a rectangular diagonal matrix with the decreasing non-negative singular values $\mathbf{\sigma}_i$ on the diagonal, i.e., $\sigma_1 \ge \sigma_i \ge \sigma_{\ell}$. $\mathbf{\Psi}$ and $\mathbf{W}$ contain the left and right singular vectors, respectively. The left singular vectors $\boldsymbol{\psi}_i$ can be interpreted as the modes of the system and the singular values $\mathbf{\sigma}_i$ as the energy content of the corresponding mode. If a steep decay of the singular values is observed, it can be assumed that the system can be well approximated by a reduced basis, using only the first $m$ modes that capture the dominant behavior of the system. The reduced basis is then given by truncating the left singular vectors, resulting in
\begin{equation}
    \mathbf{\Phi} = [\boldsymbol{\psi}_1 \, \boldsymbol{\psi}_2 \, \dots \, \boldsymbol{\psi}_m] \in \mathbb{R}^{n \times m}.
\end{equation}
In the works of \citet{kehls_reduced_2023} and \citet{zhang_multi-field_2025} it was shown that a POD projection matrix created from the unmodified snapshot matrix $\mathbf{D}$ can lead to a reduced order model (ROM) that is not necessarily stable during simulation, especially in highly nonlinear cases involving damage and plasticity. To overcome this issue, it was suggested in \citet{zhang_multi-field_2025} to decompose the snapshot matrix into separate snapshot matrices for the different field quantities, i.e. in the present context into one for the displacements $\mathbf{D}_U$ and one for the non-local damage $\mathbf{D}_{\bar{D}}$, such that
\begin{equation}
    \mathbf{D} = \mathbf{D}_U + \mathbf{D}_{\bar{D}},
\end{equation}
where
\begin{equation}
    \mathbf{D}_U =
    \begin{bmatrix}
        U^1_{1x} & \dots & U^{\ell}_{1x} \\ U^1_{1y} & \dots & U^{\ell}_{1y} \\ U^1_{1z} & \dots & U^{\ell}_{1z}\\ 0 & \dots & 0 \\ \vdots & \ddots & \vdots \\ U^1_{n_k x} & \dots & U^{\ell}_{n_k x} \\ U^1_{n_k y} & \dots & U^{\ell}_{n_k y} \\ U^1_{n_k z} & \dots & U^{\ell}_{n_k z} \\ 0 & \dots & 0
    \end{bmatrix} \in \mathbb{R}^{n \times \ell}
    , \quad
    \mathbf{D}_{\bar{D}} =
    \begin{bmatrix}
        0 & \dots & 0 \\ 0 & \dots & 0 \\ 0 & \dots & 0\\ \bar{D}^1_{1} & \dots & \bar{D}^{\ell}_{1} \\ \vdots & \ddots & \vdots \\ 0 & \dots & 0 \\ 0 & \dots & 0 \\ 0 & \dots & 0 \\ \bar{D}^1_{n_k} & \dots & \bar{D}^{\ell}_{n_k}
    \end{bmatrix} \in \mathbb{R}^{n \times \ell}.
\end{equation}

After the SVD of both snapshot matrices, the left singular vectors are truncated after $m_U$ and $m_{\bar{D}}$ modes, respectively, resulting in two projection matrices $\mathbf{\Phi}_U = [\boldsymbol{\phi}_{U,1} \, \boldsymbol{\phi}_{U,2} \, \dots \, \boldsymbol{\phi}_{U,m_U}] \in \mathbb{R}^{n \times m_U}$ for the displacements and $\mathbf{\Phi}_{\bar{D}} = [\boldsymbol{\phi}_{\bar{D},1} \, \boldsymbol{\phi}_{\bar{D},2} \, \dots \, \boldsymbol{\phi}_{\bar{D},m_{\bar{D}}}] \in \mathbb{R}^{n \times m_{\bar{D}}}$ for the non-local damage. Combining the two projection matrices results in a final projection matrix
\begin{equation}
    \mathbf{\Phi} = [\boldsymbol{\phi}_{U,1} \, \boldsymbol{\phi}_{U,2} \, \dots \, \boldsymbol{\phi}_{U,m_U} \, \boldsymbol{\phi}_{\bar{D},1} \, \boldsymbol{\phi}_{\bar{D},2} \, \dots \, \boldsymbol{\phi}_{\bar{D},m_{\bar{D}}}] \in \mathbb{R}^{n \times (m_U + m_{\bar{D}})}
\end{equation}
that can be applied as presented in \Cref{sec: pMOR}. It should be noted that the number of modes $m_U$ and $m_{\bar{D}}$ can be chosen independently, allowing for an efficient reduction of the system. The above described methodology can be applied analogously if more fields are present in the system, e.g., a temperature field. As before, the POD basis can then be constructed for each field separately and combined to form a final projection matrix.

\subsection{(Decomposed) discrete empirical interpolation method}
\label{sec: DDEIM}
The discrete empirical interpolation method (DEIM) is a hyper-reduction technique that is used to interpolate the nonlinear terms of a system. The DEIM is used to reduce the computational cost of the system by interpolating the nonlinear terms of the system with a reduced set of interpolation points. This translates to fewer element evaluations during the reduced order simulation.
To apply the DEIM to the aforementioned problem, a split into linear and nonlinear terms is performed beforehand, such that the residual vector $\boldsymbol{G}$ is given by
\begin{equation}
    \boldsymbol{G}(\boldsymbol{V}) \coloneqq \mathbf{K}_{lin} \, \boldsymbol{V} + \boldsymbol{R}_{nl}(\boldsymbol{V}) - \boldsymbol{P}.
\end{equation}
The linear term is approximated by multiplying the linear stiffness matrix $\mathbf{K}_{lin}$  -- obtained by evaluating the stiffness matrix at the zero state $\mathbf{K}_{lin} \coloneqq \mathbf{K}(\boldsymbol{0})$ -- from the right with the solution vector $\boldsymbol{V}$. The nonlinear term $\boldsymbol{R}_{nl}(\boldsymbol{V})$ is given by the remainder $\boldsymbol{R}_{nl}(\boldsymbol{V}) = \boldsymbol{R}(\boldsymbol{V}) - \mathbf{K}_{lin} \, \boldsymbol{V}$. As before, the solution vector is approximated and the Galerkin projection is applied, resulting in the reduced residual vector
\begin{equation}
    \hat{\boldsymbol{G}}(\mathbf{\Phi} \hat{\boldsymbol{V}}) \coloneqq \mathbf{\Phi}^T \, \boldsymbol{G}(\mathbf{\Phi} \hat{\boldsymbol{V}}) = \underbrace{\coloneqq \mathbf{\Phi}^T \, \mathbf{K}_{lin} \, \mathbf{\Phi}}_{\hat{\mathbf{K}}_{lin}} \hat{\boldsymbol{V}} + \mathbf{\Phi}^T \, \boldsymbol{R}_{nl}(\mathbf{\Phi} \hat{\boldsymbol{V}}) - \mathbf{\Phi}^T \, \boldsymbol{P}.
\end{equation}
The reduced linear stiffness matrix $\hat{\mathbf{K}}_{lin}$ stays constant during the simulation and needs to be computed only once at the beginning of the reduced simulation. The nonlinear term still depends on the projected solution vector $\mathbf{\Phi} \,\hat{\boldsymbol{V}}$ and, as such, the nonlinear term has to be recomputed in every iteration of the reduced iterative solution scheme. If complex material models are used, such as the one described in \Cref{sec:mat}, the computation of the residual vector and stiffness matrix can become a severe bottleneck in the simulation. To reduce the computational cost of the nonlinear term, the DEIM is applied to $\boldsymbol{R}_{nl}(\boldsymbol{V})$. For this, it is assumed that the nonlinear internal forces can also be approximated by a projection matrix $\mathbf{\Omega} \in \mathbb{R}^{n \times k}$, such that
\begin{equation}
    \label{eq:Rnlapprox}
    \boldsymbol{R}_{nl}(\boldsymbol{V}) \approx \mathbf{\Omega} \hat{\boldsymbol{R}}_{nl}(\boldsymbol{V}),
\end{equation}
where $\hat{\boldsymbol{R}}_{nl}$ is the reduced nonlinear internal force vector. The projection matrix $\mathbf{\Omega}$ is constructed using a snapshot matrix $\mathbf{D}_{nl} = [\boldsymbol{R}_{nl}(\boldsymbol{V}_1) \, \boldsymbol{R}_{nl}(\boldsymbol{V}_2) \, \dots \, \boldsymbol{R}_{nl}(\boldsymbol{V}_{\ell})] \in \mathbb{R}^{n \times \ell}$ containing snapshots of the nonlinear part of the internal force vector. Afterwards, a SVD of the snapshot matrix is performed, resulting in
\begin{equation}
    \mathbf{D}_{nl} = \mathbf{\Psi}_{nl} \mathbf{\Sigma}_{nl} \mathbf{W}_{nl}.
\end{equation}
The matrix $\mathbf{\Psi}_{nl}$ containing the left singular vectors $\boldsymbol{\psi}_{nl,i}$ is then truncated after $k$ modes, resulting in the projection matrix
\begin{equation}
    \mathbf{\Omega} = [\boldsymbol{\psi}_{nl,1} \, \boldsymbol{\psi}_{nl,2} \, \dots \, \boldsymbol{\psi}_{nl,k}] \in \mathbb{R}^{n \times k}.
\end{equation}
To overcome the dependence of the residual vector and tangential stiffness matrix on the solution vector, a selection matrix $\mathbf{Z} = [\boldsymbol{e}_{p_1} \, \boldsymbol{e}_{p_2} \, \dots \, \boldsymbol{e}_{p_k}] \in \mathbb{R}^{n \times k}$ is introduced, which contains $k$ unit vectors $\boldsymbol{e}_{p_i}$ corresponding to the interpolation DOFs $p_i$. The multiplication of the transposed selection matrix $\mathbf{Z}^T$ with \autoref{eq:Rnlapprox} leads to the following equation
\begin{equation}
    \mathbf{Z}^T \, \boldsymbol{R}_{nl}(\boldsymbol{V}) \approx \mathbf{Z}^T \, \mathbf{\Omega} \hat{\boldsymbol{R}}_{nl}(\boldsymbol{V}).
\end{equation}
Reformulation of the above equation leads to the following expression for the reduced nonlinear internal force vector
\begin{equation}
    \hat{\boldsymbol{R}}_{nl}(\boldsymbol{V}) = (\mathbf{Z}^T \, \mathbf{\Omega})^{-1} \, \mathbf{Z}^T \, \boldsymbol{R}_{nl}(\boldsymbol{V}).
\end{equation}
Using this relation, the approximation of the nonlinear internal force vector can be reformulated as
\begin{equation}
    \boldsymbol{R}_{nl}(\boldsymbol{V}) \approx \mathbf{\Omega} \, (\mathbf{Z}^T \, \mathbf{\Omega})^{-1} \, \mathbf{Z}^T \, \boldsymbol{R}_{nl}(\boldsymbol{V}).
\end{equation}
Finally, the reduced residual vector can be expressed as
\begin{equation}
    \begin{aligned}
        \hat{\boldsymbol{G}}(\mathbf{\Phi} \hat{\boldsymbol{V}}) \coloneqq & \, \hat{\mathbf{K}}_{lin} \, \hat{\boldsymbol{V}} + \mathbf{\Phi}^T \, \mathbf{\Omega} \, (\mathbf{Z}^T \, \mathbf{\Omega})^{-1} \, \mathbf{Z}^T \boldsymbol{R}_{nl}(\mathbf{\Phi} \hat{\boldsymbol{V}}) - \mathbf{\Phi}^T \, \boldsymbol{P} \\
        =                                                                  & \, \hat{\mathbf{K}}_{lin} \, \hat{\boldsymbol{V}} + \mathbf{M}_{\textrm{DEIM}} \mathbf{Z}^T \boldsymbol{R}_{nl}(\mathbf{\Phi} \hat{\boldsymbol{V}}) - \mathbf{\Phi}^T \, \boldsymbol{P}.
    \end{aligned}
\end{equation}
The newly introduced matrix $\mathbf{M}_{\textrm{DEIM}} \coloneqq \mathbf{\Phi}^T \, \mathbf{\Omega} \, (\mathbf{Z}^T \, \mathbf{\Omega})^{-1} \in \mathbb{R}^{m \times k}$ is a constant matrix that needs to be computed only once in the beginning of the reduced simulation. The reduced iterative solution scheme is then given by
\begin{equation}
    \label{eq:DEIMsoe}
    \begin{aligned}
         & (\hat{\mathbf{K}}_{lin} + \mathbf{M}_{\textrm{DEIM}} \mathbf{Z}^T \mathbf{K}_{nl}(\mathbf{\Phi} \hat{\boldsymbol{V}}^i) \mathbf{\Phi}) \, \Delta \hat{\boldsymbol{V}}^{i+1} = - \hat{\boldsymbol{G}}(\mathbf{\Phi} \hat{\boldsymbol{V}}^i) \\
         & \boldsymbol{V}^{i+1} = \boldsymbol{V}^{i} + \mathbf{\Phi} \Delta \hat{\boldsymbol{V}}^{i+1}                                                                                                                                                \\
         & ||\hat{\boldsymbol{G}}(\mathbf{\Phi} \hat{\boldsymbol{V}}^i)|| < tol                                                                                                                                                                       \\
         & i = i + 1.
    \end{aligned}
\end{equation}
The multiplication of the selection matrix $\mathbf{Z}^T$ with the nonlinear internal force vector $\boldsymbol{R}_{nl}(\mathbf{\Phi} \hat{\boldsymbol{V}})$ as well as the nonlinear part of the stiffness matrix $\mathbf{K}_{nl}(\mathbf{\Phi} \hat{\boldsymbol{V}})$ leads in practice to a reduced number of element evaluations, as only elements containing an interpolation DOF $p_i$ have to be evaluated. The multiplication of the selection matrix $\mathbf{Z}^T$ can be implemented as a slicing operation, which is computationally more efficient than a matrix multiplication.

To find suitable interpolation points, the greedy algorithm based on \cite{chaturantabut_nonlinear_2010} shown in \Cref{algo:DEIM} is used.
\begin{algorithm}[ht]
    \caption{Computation of DEIM interpolation DOFs based on \cite{chaturantabut_nonlinear_2010}.}
    \label{algo:DEIM}
    \KwIn{$\mathbf{\Omega} \in \mathbb{R}^{n \times k}$}
    \KwOut{$\mathbf{Z} \in \mathbb{R}^{n \times k}$}
    $\gamma_1 = \text{argmax}(|\boldsymbol{\phi}_{nl,1}|)$\\
    $\mathbf{Z}_1 = [\boldsymbol{e}_{\gamma_1}]$\\
    $\mathbf{\Omega}_1 = [\boldsymbol{\phi}_{nl,1}]$\\
    $i = 1$\\
    \While{$i < k$}{
    $\boldsymbol{c}_{i+1} = (\mathbf{Z}_i^T \mathbf{\Omega}_i)^{-1} \mathbf{Z}_i^T \boldsymbol{\phi}_{nl,i+1}$\\
    $\boldsymbol{r}_{i+1} = \boldsymbol{\phi}_{nl,i+1} - \mathbf{\Omega}_i \boldsymbol{c}_{i+1}$\\
    $\gamma_{i+1} = \text{argmax}(|\boldsymbol{r}_{i+1}|)$\\
    $\mathbf{Z}_{i+1} = [\mathbf{Z}_i \, \boldsymbol{e}_{\gamma_{i+1}}]$\\
    $\mathbf{\Omega}_{i+1} = [\mathbf{\Omega}_i \, \boldsymbol{\phi}_{nl,i+1}]$\\
    $i = i + 1$
    }
\end{algorithm}
The \textit{argmax} function returns the index of the maximum value of the vector and the vector $\boldsymbol{e}_{\gamma_i}$ is a unit vector with a $1$ at the index $\gamma_i$ and $0$ elsewhere. For further details and an illustration of the greedy algorithm, the reader is referred to \citet{chaturantabut_nonlinear_2010}.

To extend the DEIM for multi-field problems, the decomposed projection matrix described in \Cref{sec: DPOD} is used again. In addition to that, the nonlinear internal force vector $\boldsymbol{R}_{nl}(\boldsymbol{V})$ is decomposed into a displacement and a non-local damage part, analogously to the snapshots of the solution vector, such that
\begin{equation}
    \mathbf{D}_{nl} = \mathbf{D}_{nl,U} + \mathbf{D}_{nl,\bar{D}},
\end{equation}
where $\mathbf{D}_{nl,U}$ contains the snapshots of the nonlinear internal force vector connected to the displacements and $\mathbf{D}_{nl,\bar{D}}$ the snapshots of the nonlinear internal force vector connected to non-local damage. Applying the SVD to both parts separately, results in two projection matrices $\mathbf{\Omega}_U \in \mathbb{R}^{n \times k_U}$ and $\mathbf{\Omega}_{\bar{D}} \in \mathbb{R}^{n \times k_{\bar{D}}}$, where $k_U$ and $k_{\bar{D}}$ are the number of interpolation points for the displacements and the non-local damage, respectively. The selection matrix is then computed for each projection matrix of the nonlinear internal forces individually, using \Cref{algo:DEIM} two times, resulting in two selection matrices $\mathbf{Z}_U \in \mathbb{R}^{n \times k_U}$ and $\mathbf{Z}_{\bar{D}} \in \mathbb{R}^{n \times k_{\bar{D}}}$. The final selection matrix is then given by
\begin{equation}
    \mathbf{Z} = [\mathbf{Z}_U \, \mathbf{Z}_{\bar{D}}] \in \mathbb{R}^{n \times (k_U + k_{\bar{D}})}
\end{equation}
and the final projection matrix by
\begin{equation}
    \mathbf{\Omega} = [\mathbf{\Omega}_U \, \mathbf{\Omega}_{\bar{D}}] \in \mathbb{R}^{n \times (k_U + k_{\bar{D}})}.
\end{equation}
These two matrices can be used in a straight forward manner in the DEIM-reduced iterative solution scheme, described in \autoref{eq:DEIMsoe}. It should be noted that the number of interpolation points $k_U$ and $k_{\bar{D}}$ can be chosen independently. It is not directly possible to calculate the reduced number of element evaluations $n_{\hat{\mathcal{E}}_k}$ from the number of interpolation DOFs, because in the three-dimensional case up to four interpolation DOFs can belong to the same node, leading to no additional element evaluations. In general, however, the number of necessary element evaluations increases - and therefore the computation cost savings decrease -  with an increasing number of interpolation DOFs.

\subsection{(Decomposed) energy conserving sampling and weighting method}
\label{sec: ECSWM}
Similar to the DEIM, the energy conserving sampling and weighting method is a POD-based hyper-reduction technique that seeks to reduce the computational cost of the system by reducing the number of element evaluations. Instead of interpolating the nonlinear internal force vector and the nonlinear stiffness matrix first and then projecting them to the reduced solution space, the ECSW directly approximates the reduced nonlinear internal force vector and the reduced nonlinear stiffness matrix. To achieve this, the reduced internal force vector is rewritten as a sum over the original element set $\mathcal{E}$, such that
\begin{equation}
    \hat{\boldsymbol{R}}(\boldsymbol{V}) = \mathbf{\Phi}^T \, \boldsymbol{R}(\boldsymbol{V}) = \sum_{e \in \mathcal{E}} \mathbf{\Phi}_e^T \, \boldsymbol{R}_e(\boldsymbol{V}_e),
\end{equation}
where the subscript $e$ indicates that only the values corresponding to the DOFs of the element $e$ are considered. The goal of the ECSW is to approximate the reduced internal force vector $\hat{\boldsymbol{R}}(\boldsymbol{V})$ using a reduced set of elements $\hat{\mathcal{E}} \subset \mathcal{E}$ and corresponding positive element weights $w_e > 0$, such that
\begin{equation}
    \label{eq:ECSWM}
    \hat{\boldsymbol{R}}(\boldsymbol{V}) = \sum_{e \in \mathcal{E}} \mathbf{\Phi}_e^T \, \boldsymbol{R}_e(\boldsymbol{V}_e) \approx \sum_{e \in \hat{\mathcal{E}}} w_e \mathbf{\Phi}_e^T \, \boldsymbol{R}_e(\boldsymbol{V}_e).
\end{equation}
For later purposes, \autoref{eq:ECSWM} is rewritten in matrix notation and for all $\ell$ snapshots, such that
\begin{equation}
    \underbrace{
        \begin{bmatrix}
            \mathbf{\Phi}_1^T \boldsymbol{R}_1 (\mathbf{\Omega} \boldsymbol{V}_{1, 1}) & \dots  & \mathbf{\Phi}_{n_{\mathcal{E}}}^T \boldsymbol{R}_{n_{\mathcal{E}}} (\mathbf{\Omega} \boldsymbol{V}_{n_{\mathcal{E}},1})     \\
            \vdots                                                                     & \ddots & \vdots                                                                                                                      \\
            \mathbf{\Phi}_1^T \boldsymbol{R}_1 (\mathbf{\Omega} \boldsymbol{V}_{\ell}) & \dots  & \mathbf{\Phi}_{n_{\mathcal{E}}}^T \boldsymbol{R}_{n_{\mathcal{E}}} (\mathbf{\Omega} \boldsymbol{V}_{n_{\mathcal{E}}, \ell}) \\
        \end{bmatrix}}_{\coloneqq \mathbf{Y}}
    \underbrace{
        \begin{bmatrix}
            w_1 \\ \vdots \\ w_{n_{\mathcal{E}}}
        \end{bmatrix}}_{\coloneqq \boldsymbol{w}}
    =
    \underbrace{
        \begin{bmatrix}
            \sum_{e\in \mathcal{E}} \mathbf{\Phi}_e^T \boldsymbol{R}_e (\mathbf{\Omega} \boldsymbol{V}_1) \\ \vdots \\
            \sum_{e\in \mathcal{E}} \mathbf{\Phi}_e^T \boldsymbol{R}_e (\mathbf{\Omega} \boldsymbol{V}_{\ell})
        \end{bmatrix}}_{\coloneqq \boldsymbol{b}}.
\end{equation}
The above equation is fulfilled if all weights $w_e$ are set to $1$. To reduce the number of element evaluations, a weight vector $\boldsymbol{w}$ with as many zero entries as possible is computed, approximately fulfilling \autoref{eq:ECSWM}. Once the feasibility set for the weights is defined as
\begin{equation}
    \mathcal{P} = \{\boldsymbol{w} : ||\mathbf{Y} \boldsymbol{w} - \boldsymbol{b}||_2 < \tau || \boldsymbol{b}||_2, \, w_e \ge 0\},
\end{equation}
where $\tau \ge 0$ is a small user-chosen tolerance, the optimal weight vector $\boldsymbol{w}^{*}$ is defined as the solution of the constrained optimization problem
\begin{equation}
    \boldsymbol{w}^{*} = \text{arg} \, \underset{w \in \mathcal{P}}{\text{min}} \, || \boldsymbol{w} ||_0,
\end{equation}
where $|| \bullet ||_0$ denotes the $l_0$-norm, which yields the number of non-zero entries in the vector $\bullet$.
Because the constrained optimization problem can not be solved directly, it is substituted by the inexact non-negative least-squares problem
\begin{equation}
    \boldsymbol{w}^{*} \approx \text{arg} \, \underset{w \in \mathcal{Q}}{\text{min}} \,||\mathbf{Y} \boldsymbol{w} - \boldsymbol{b}||_2, \quad \mathcal{Q} =  \{\boldsymbol{w} : w_e \ge 0\}.
\end{equation}
The problem is solved for $\boldsymbol{w}$ using a sparse non-negative least-squares solver (sNNLS) until the condition
\begin{equation}
    ||\mathbf{Y} \boldsymbol{w} - \boldsymbol{b}||_2 < \tau || \boldsymbol{b}||_2
\end{equation}
is met.
Having computed a weight vector $\boldsymbol{w}^{*}$ that fulfills the above condition, the reduced internal force vector can be computed as
\begin{equation}
    \hat{\boldsymbol{R}}(\boldsymbol{V}) = \sum_{e \in \hat{\mathcal{E}}} w_e \mathbf{\Phi}_e^T \, \boldsymbol{R}_e(\boldsymbol{V}_e).
\end{equation}
The reduced tangential stiffness matrix can be computed analogously, such that
\begin{equation}
    \hat{\mathbf{K}}_T(\boldsymbol{V}) = \sum_{e \in \hat{\mathcal{E}}} w_e \mathbf{\Phi}_e^T \, \mathbf{K}_{e}(\boldsymbol{V}_e) \, \mathbf{\Phi}_e.
\end{equation}
Both terms can then be used in the reduced iteration solution scheme given in \autoref{eq:reducedsoe}. It is interesting to note that no special treatment is necessary to apply the decomposition with regards to different solutions fields to the ECSW. The decomposition comes solely from the decomposition of the POD projection matrix $\mathbf{\Phi}$, which is used during the computation of the weights in the offline phase as well as the projection of the nonlinear internal force vector and the tangential stiffness matrix in the online phase.

\section{Numerical examples}
\label{sec: numerical examples}
In the following, we present numerical examples to investigate the performance of the different proposed methods. Two specimens are used to thoroughly test the accuracy of the methods, while a third example is used to demonstrate the applicability of the methods to an optimization problem. Unless otherwise specified, the material model from \Cref{sec:mat} with the material parameters from \autoref{tab: material parameters} are used.
\begin{table}[htbp]
    \centering
    \caption{Parameter set for the numerical examples.}
    \label{tab: material parameters}
    \begin{tabular}{l  l  l  l}
        Symbol     & Material parameter          \hspace{4cm} & Value \hspace{1.5cm} & Unit                     \\
        \hline
        $\Lambda$  & first Lamé parameter                     & 25000                & MPa                      \\
        $\mu$      & second Lamé parameter                    & 55000                & MPa                      \\
        $\sigma_0$ & yield stress                             & 400                  & MPa                      \\
        $a$        & first kinematic hardening parameter      & 450                  & MPa                      \\
        $e$        & first isotropic hardening parameter      & 265                  & MPa                      \\
        $f$        & second isotropic hardening parameter     & 16.93                & [-]                      \\
        $Y_0$      & damage threshold                         & 2.5                  & MPa                      \\
        $r$        & first damage hardening parameter         & 5                    & MPa                      \\
        $s$        & second damage hardening parameter        & 10                   & MPa                      \\
        $A$        & internal length scale parameter          & 500                  & MPa $\text{mm}^\text{2}$ \\
        $H$        & penalty parameter                        & $\text{10}^\text{4}$ & MPa
    \end{tabular}
\end{table}
\subsection{Smiley specimen}
\begin{figure}[htbp]
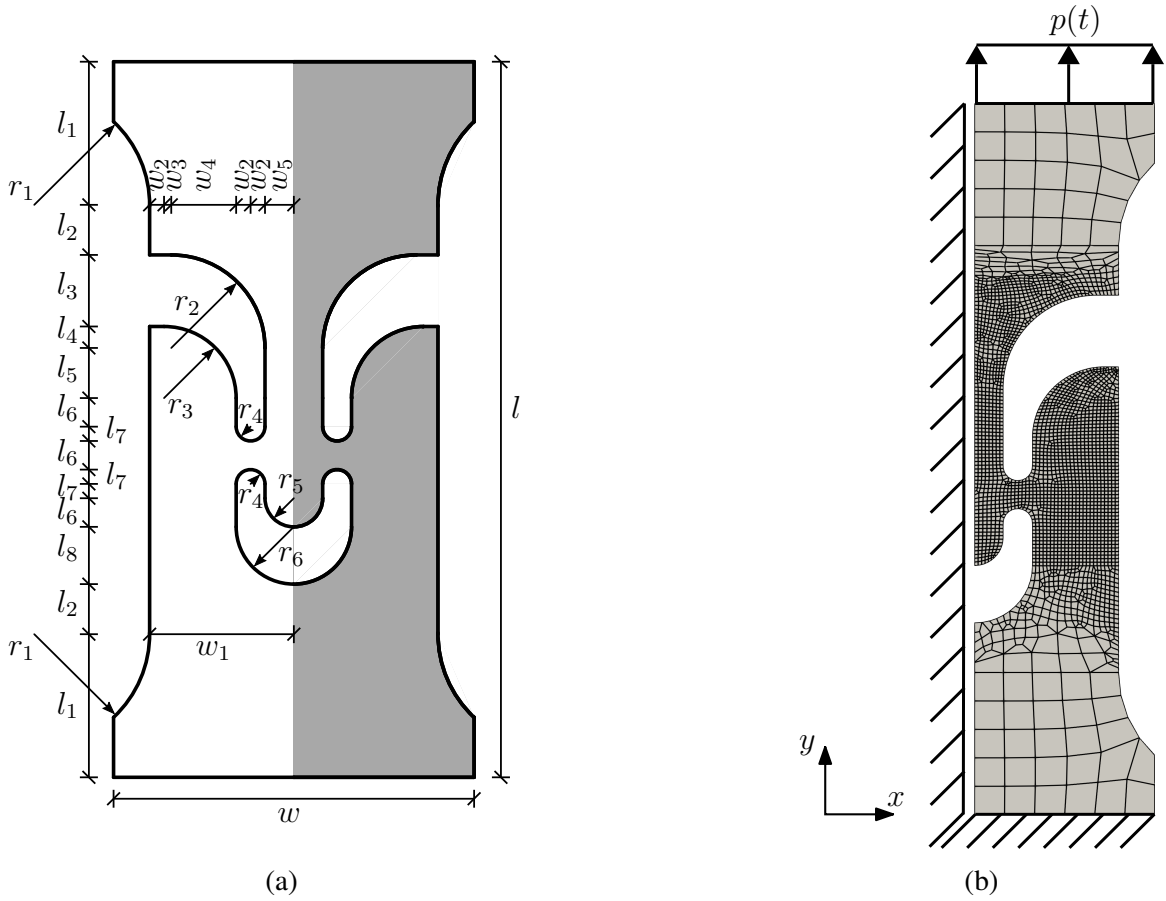

    \centering
    \begin{subfigure}[t]{0.45\textwidth}
        \centering
        \includesvg[width=\textwidth]{images/geom_smiley.svg}
        \caption{}
        \label{fig: smiley specimen}
    \end{subfigure}
    \hspace{3cm}
    \begin{subfigure}[t]{0.3\textwidth}
        \centering
        \includesvg[width=\textwidth]{images/smiley_sym.svg}
        \caption{}
        \label{fig: smiley specimen sym}
    \end{subfigure}
    \caption{(a) Original geometry of the \textit{smiley specimen} with dimensions $l = 50$ mm, $l_1 = 10$ mm, $l_2 = 3.5$ mm, $l_3 = 5$ mm, $l_4 = 1.5$ mm, $l_5 = 4.5$ mm, $l_6 = 2$ mm, $l_7 = 1$ mm, $l_8 = 4$ mm, $w = 25$ mm, $w_1 = 10$ mm, $w_2 = 1$ mm, $w_3 = 0.5$ mm, $w_4 = 4.5$ mm, $w_5 = 2$ mm, $r_1 = 8$ mm, $r_2 = 6.5$ mm, $r_3 = 5$ mm, $r_4 = 1$ mm, $r_5 = 2$ mm, and $r_6 = 4$ mm and a thickness of $1$ mm. (b) Discretization into 2649 linear quadrilateral finite elements and boundary conditions (symmetry exploited).}
\end{figure}

\begin{figure}[htbp]
    \begin{subfigure}[b]{0.65\textwidth}
        \centering
        \pgfplotsset{%
            width=\textwidth,
            height=0.7\textwidth
        }
        \input{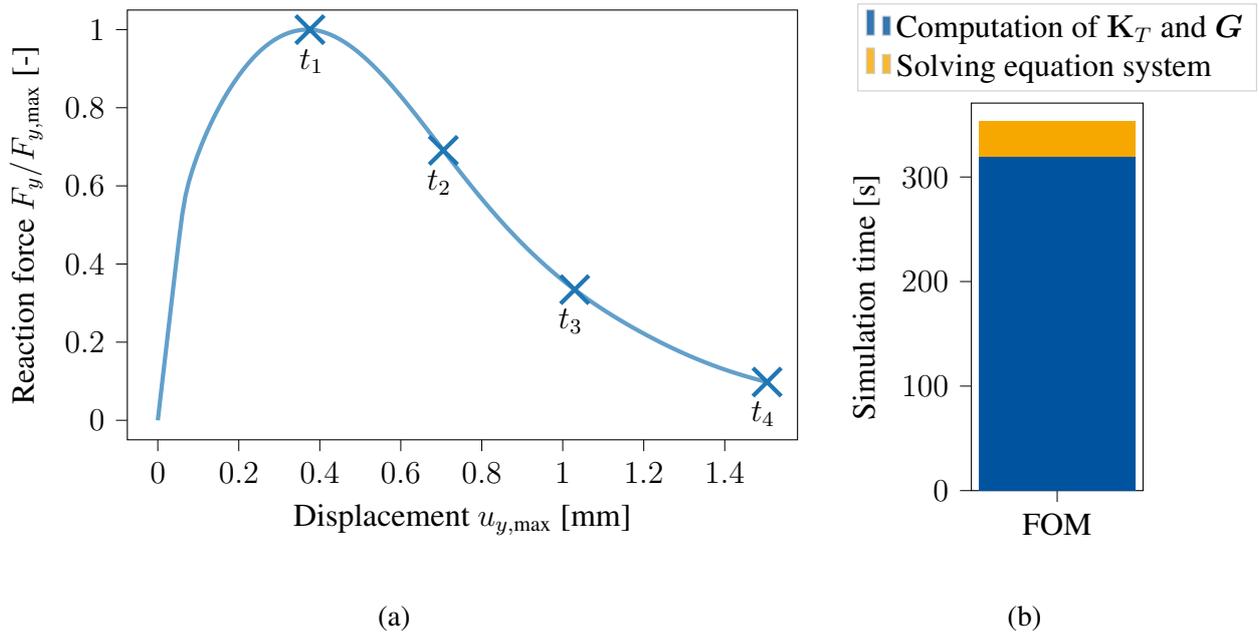}
        \caption{}
        \label{fig: force disp}
    \end{subfigure}
    \hfill
    \begin{subfigure}[b]{0.3\textwidth}
        \centering
        \pgfplotsset{%
            width=0.8\textwidth,
            height=1.4\textwidth
        }
\begin{tikzpicture}

  \definecolor{darkgray176}{RGB}{176,176,176}
  \definecolor{lightgray204}{RGB}{204,204,204}
  \definecolor{orange2461680}{RGB}{246,168,0}
  \definecolor{teal084159}{RGB}{0,84,159}

  \begin{axis}[
      legend cell align={left},
      legend style={
          fill opacity=0.8,
          draw opacity=1,
          text opacity=1,
          at={(0.5,1.03)},
          anchor=south,
          draw=lightgray204
        },
      tick align=outside,
      tick pos=left,
      x grid style={darkgray176},
      xmin=-0.44, xmax=0.44,
      xtick style={color=black},
      xtick={0},
      xticklabels={FOM},
      y grid style={darkgray176},
      ylabel={Simulation time [s]},
      ymin=0, ymax=370.686218369007,
      ytick style={color=black}
    ]
    \draw[draw=none,fill=teal084159] (axis cs:-0.4,0) rectangle (axis cs:0.4,319.291780948639);
    \addlegendimage{ybar,ybar legend,draw=none,fill=teal084159}
    \addlegendentry{Computation of $\mathbf{K}_T$ and $\boldsymbol{G}$}

    \draw[draw=none,fill=orange2461680] (axis cs:-0.4,319.291780948639) rectangle (axis cs:0.4,353.034493684769);
    \addlegendimage{ybar,ybar legend,draw=none,fill=orange2461680}
    \addlegendentry{Solving equation system}

  \end{axis}

\end{tikzpicture}
        \caption{}
        \label{fig: time}
    \end{subfigure}
    \label{fig: force disp time}
    \caption{(a) Force-displacement curve of the \textit{smiley specimen}. Four pseudo-timesteps $t_1$, $t_2$, $t_3$, and $t_4$ are marked for reference. (b) Simulation time for the full-order simulation of the \textit{smiley specimen} broken down into time spent on the computation of the tangential stiffness matrix $\mathbf{K}_T$ as well as residual vector $\boldsymbol{G}$ and the time spent solving the equation system.}
\end{figure}

\begin{figure}[htbp]
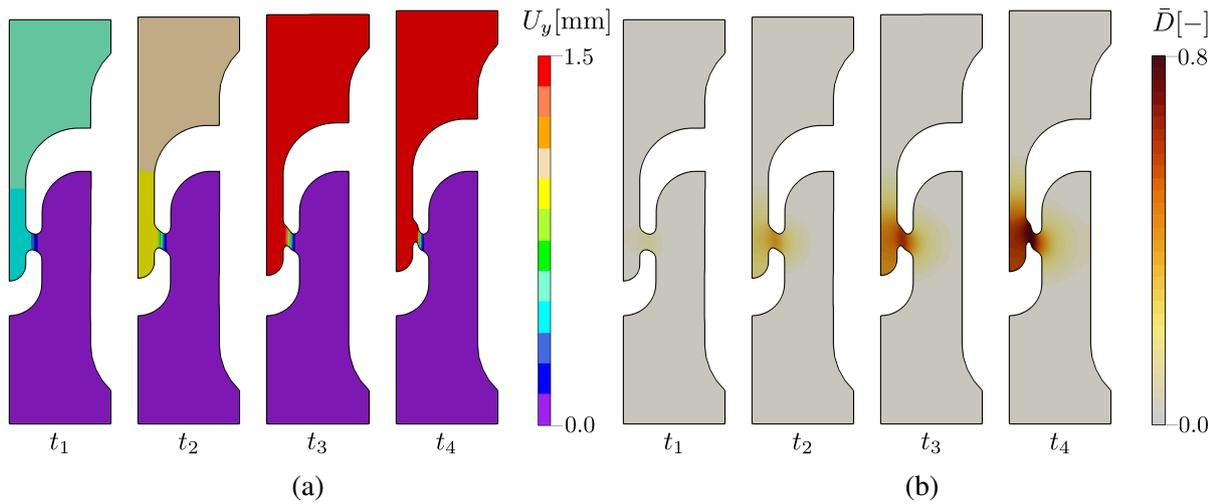

    \begin{subfigure}[t]{0.5\textwidth}
        \centering
        \includesvg[width=\textwidth]{images/U_y.svg}
        \caption{}
        \label{fig: U_y}
    \end{subfigure}
    \begin{subfigure}[t]{0.5\textwidth}
        \centering
        \includesvg[width=\textwidth]{images/Dbar.svg}
        \caption{}
        \label{fig: Dbar}
    \end{subfigure}
    \caption{Contour plots of the displacement in $y$-direction and the non-local damage variable at four pseudo-timesteps $t_1$, $t_2$, $t_3$, and $t_4$.}
\end{figure}

The first example is the so-called \textit{smiley specimen}, which is a two-dimensional plane-strain specimen taken from \cite{van_der_velden_comparative_2024}. The boundary value problem is shown in \autoref{fig: smiley specimen}. The bottom edge is fixed in $x$- and $y$-direction and a force is applied incrementally at the top edge. Due to the symmetry of the problem, only half of the specimen is considered. The discretization into 2649 bilinear quadrilateral finite elements and the final boundary conditions are shown in \autoref{fig: smiley specimen sym}. The discretization was chosen on the basis of a mesh convergence study, which showed that no further refinement was necessary. All simulations are performed using the arc-length method. This enables the simulations to continue deep into the softening regime, even though the simulations are force-controlled. The corresponding force-displacement curve is shown in \autoref{fig: force disp}. Additionally, the simulation time is broken down into the time to compute the tangential stiffness matrix $\mathbf{K}_T$ and internal force vector $\boldsymbol{G}$ and the time to solve the equation system. Due to the complex material model, the simulation time is dominated by the computation of the tangential stiffness matrix and internal force vector. Contour plots of the displacement in $y$-direction and the non-local damage variable are shown for four different pseudo-timesteps $t_1$, $t_2$, $t_3$, and $t_4$ in \autoref{fig: U_y} and \autoref{fig: Dbar}, respectively. The four pseudo-timesteps are also marked in \autoref{fig: force disp}. It can be seen nicely, that the used gradient-extended damage-plasticity formulation leads to a ductile failure behavior of the specimen with a reasonably looking, finite damage zone in the expected critical area of the specimen. No spurious mesh-sensitivity issues are otherwise observed.

\subsubsection*{DPOD}
The first investigation is concerned with the decomposed POD method as this indicates the potential of the hyper-reduction approaches, which are based on the POD. The POD basis is computed using the snapshots from the full-order simulation shown in \autoref{fig: force disp}. In total, $\ell = 219$ snapshots are used. The same problem as before is simulated with the decomposed POD method and the influence of the number of displacement modes $m_U$ and damage modes $m_D$ on the accuracy is investigated. Due to use of the arc-length method, the pseudo-timesteps of the full-order simulation and the reduced-order simulation do generally not share the same displacement or force. To compute the relative error $\epsilon$, $N=1000$ points are sampled from the force-displacement curves that share the same displacement. The error is then computed as the mean error of the reaction force corresponding to the sampled points such that
\begin{equation}
    \epsilon = \frac{1}{N} \sum_{i=1}^{N} \frac{\left| F_y^{\text{full}}(u_{y,\text{max}}^i) - F_y^{\text{reduced}}(u_{y,\text{max}}^i) \right|}{F_y^{\text{full}}(u_{y,\text{max}}^i)}.
\end{equation}
The results are shown in \autoref{fig: POD split error smiley}. A trend is observed that the error decreases with an increasing number of modes $m_U$ and $m_D$. With $m_U = 50$ and $m_D = 25$, the error is in the order of $10^{-6}$. Because the simulation time is dominated by the computation of the tangential stiffness matrix and internal force vector, the simulation time is not significantly reduced by the decomposed POD method. The simulation time ratio $\frac{T_\mathrm{DPOD}}{T_\mathrm{FOM}}$ is shown for $m_U = \{25, 50\}$ and $m_D = \{5, 10, 15, 20, 25\}$ in \autoref{fig: POD split time smiley}. Using as low as $m_U = 5$ and $m_D = 5$, the simulation time is only reduced to about 85\% of the full-order simulation time. With $m_U = 50$ and $m_D = 25$, the simulation time is reduced to about 90\% of the full-order simulation time. This shows that the decomposed POD alone is not sufficient to achieve a significant reduction of computation time in this nonlinear problem.
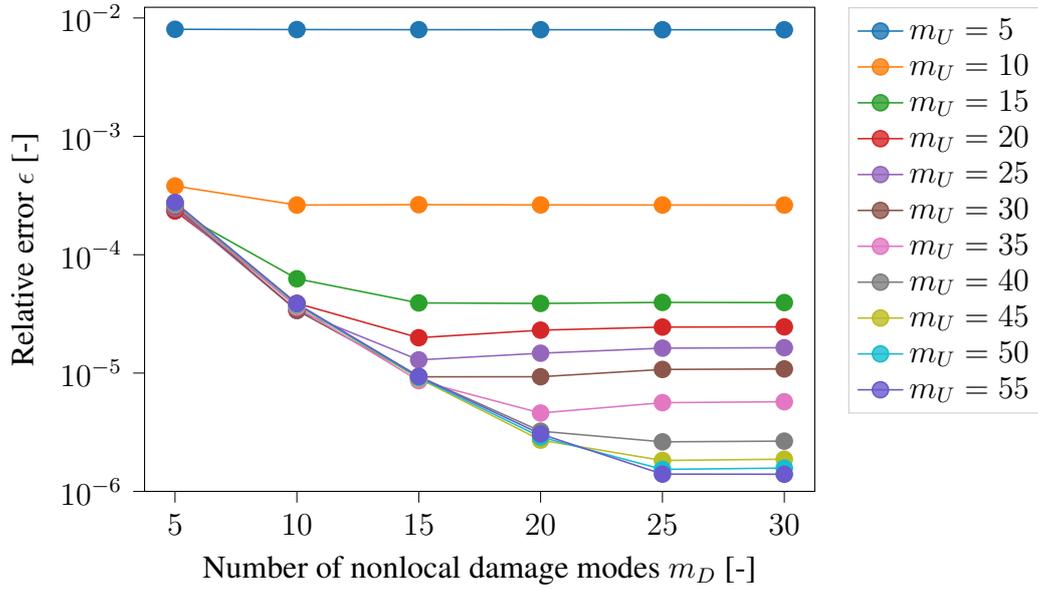
\begin{figure}[htbp]
    \centering
    \pgfplotsset{%
        width=0.65\textwidth,
        height=0.5\textwidth
    }
\begin{tikzpicture}

  \definecolor{crimson2143940}{RGB}{214,39,40}
  \definecolor{darkgray176}{RGB}{176,176,176}
  \definecolor{darkorange25512714}{RGB}{255,127,14}
  \definecolor{darkturquoise23190207}{RGB}{23,190,207}
  \definecolor{forestgreen4416044}{RGB}{44,160,44}
  \definecolor{goldenrod18818934}{RGB}{188,189,34}
  \definecolor{gray127}{RGB}{127,127,127}
  \definecolor{lightgray204}{RGB}{204,204,204}
  \definecolor{mediumpurple148103189}{RGB}{148,103,189}
  \definecolor{orchid227119194}{RGB}{227,119,194}
  \definecolor{sienna1408675}{RGB}{140,86,75}
  \definecolor{steelblue31119180}{RGB}{31,119,180}
  \definecolor{slateblue10690205}{RGB}{106,90,205}

  \begin{axis}[
      legend cell align={left},
      legend style={
          fill opacity=0.8,
          draw opacity=1,
          text opacity=1,
          at={(1.05,1)},
          anchor=north west,
          draw=lightgray204
        },
      log basis y={10},
      tick align=outside,
      tick pos=left,
      x grid style={darkgray176},
      ylabel = {Relative error $\epsilon$ [-]},
      xlabel = {Number of nonlocal damage modes $m_D$ [-]},
      xmin=3.75, xmax=31.25,
      xtick style={color=black},
      y grid style={darkgray176},
      ymin=1.00112086595609e-06, ymax=0.0123136184764605,
      ymode=log,
      ytick style={color=black},
      ytick={1e-07,1e-06,1e-05,0.0001,0.001,0.01,0.1,1},
      yticklabels={
          \(\displaystyle {10^{-7}}\),
          \(\displaystyle {10^{-6}}\),
          \(\displaystyle {10^{-5}}\),
          \(\displaystyle {10^{-4}}\),
          \(\displaystyle {10^{-3}}\),
          \(\displaystyle {10^{-2}}\),
          \(\displaystyle {10^{-1}}\),
          \(\displaystyle {10^{0}}\)
        }
    ]
    \addplot [semithick, steelblue31119180, mark=*, mark size=3, mark options={solid}]
    table {%
        5 0.00802566790696714
        10 0.00798352182554248
        15 0.00796587625544971
        20 0.00795951232469752
        25 0.00795592400092643
        30 0.00795173389966481
      };
    \addlegendentry{$m_U = 5$}
    \addplot [semithick, darkorange25512714, mark=*, mark size=3, mark options={solid}]
    table {%
        5 0.000380592758236778
        10 0.000262658762014331
        15 0.000264673517868687
        20 0.00026341034640851
        25 0.000262867466834274
        30 0.000262474047336766
      };
    \addlegendentry{$m_U = 10$}
    \addplot [semithick, forestgreen4416044, mark=*, mark size=3, mark options={solid}]
    table {%
        5 0.000236879228435476
        10 6.26350797043171e-05
        15 3.91382413546298e-05
        20 3.87701858768475e-05
        25 3.95399975284135e-05
        30 3.93995195411089e-05
      };
    \addlegendentry{$m_U = 15$}
    \addplot [semithick, crimson2143940, mark=*, mark size=3, mark options={solid}]
    table {%
        5 0.000234092944391437
        10 3.88441215820238e-05
        15 1.98980253129693e-05
        20 2.30420288025739e-05
        25 2.44657771920443e-05
        30 2.45715264515686e-05
      };
    \addlegendentry{$m_U = 20$}
    \addplot [semithick, mediumpurple148103189, mark=*, mark size=3, mark options={solid}]
    table {%
        5 0.000247124154444349
        10 3.36684233903482e-05
        15 1.29220637868772e-05
        20 1.4715547646438e-05
        25 1.62383923752291e-05
        30 1.63849398858752e-05
      };
    \addlegendentry{$m_U = 25$}
    \addplot [semithick, sienna1408675, mark=*, mark size=3, mark options={solid}]
    table {%
        5 0.000255393818843915
        10 3.38993849011929e-05
        15 9.29796104466434e-06
        20 9.31143510891181e-06
        25 1.0731341743146e-05
        30 1.0835001285704e-05
      };
    \addlegendentry{$m_U = 30$}
    \addplot [semithick, orchid227119194, mark=*, mark size=3, mark options={solid}]
    table {%
        5 0.000262655905432776
        10 3.61156841674251e-05
        15 8.62007432055376e-06
        20 4.59854749918768e-06
        25 5.62240839218767e-06
        30 5.72486725614783e-06
      };
    \addlegendentry{$m_U = 35$}
    \addplot [semithick, gray127, mark=*, mark size=3, mark options={solid}]
    table {%
        5 0.000266039714935033
        10 3.7610713986062e-05
        15 9.36816568302917e-06
        20 3.22963530402826e-06
        25 2.62439709064036e-06
        30 2.66127840593062e-06
      };
    \addlegendentry{$m_U = 40$}
    \addplot [semithick, goldenrod18818934, mark=*, mark size=3, mark options={solid}]
    table {%
        5 0.000272206014511295
        10 3.77803594760078e-05
        15 9.02207791856153e-06
        20 2.71161767660095e-06
        25 1.82600119716227e-06
        30 1.87258590308306e-06
      };
    \addlegendentry{$m_U = 45$}
    \addplot [semithick, darkturquoise23190207, mark=*, mark size=3, mark options={solid}]
    table {%
        5 0.000275394893443917
        10 3.81799844614471e-05
        15 9.16958982587809e-06
        20 2.88596856331455e-06
        25 1.53599931309212e-06
        30 1.57549068821561e-06
      };
    \addlegendentry{$m_U = 50$}
    \addplot [semithick, slateblue10690205, mark=*, mark size=3, mark options={solid}]
    table {%
        5 0.000276866180205495
        10 3.86529812505473e-05
        15 9.35523506345853e-06
        20 3.05566196304685e-06
        25 1.40087833323803e-06
        30 1.39755423804634e-06
      };
    \addlegendentry{$m_U = 55$}
  \end{axis}

\end{tikzpicture}
    \caption{Error of the DPOD simulation compared to the full-order simulation for different numbers of modes $m_U$ and $m_D$.}
    \label{fig: POD split error smiley}
\end{figure}
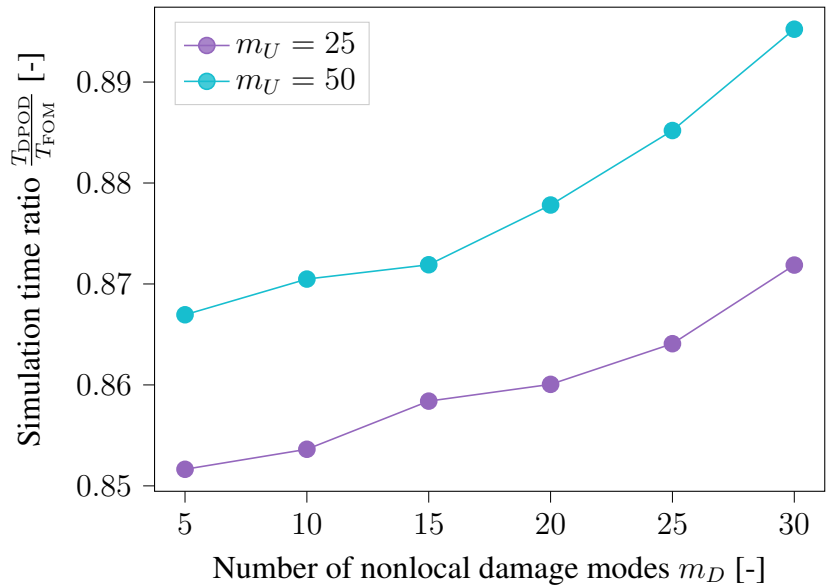
\begin{figure}[htbp]
    \centering
    \pgfplotsset{%
        width=0.65\textwidth,
        height=0.5\textwidth
    }
\begin{tikzpicture}

  \definecolor{crimson2143940}{RGB}{214,39,40}
  \definecolor{darkgray176}{RGB}{176,176,176}
  \definecolor{darkorange25512714}{RGB}{255,127,14}
  \definecolor{darkturquoise23190207}{RGB}{23,190,207}
  \definecolor{forestgreen4416044}{RGB}{44,160,44}
  \definecolor{goldenrod18818934}{RGB}{188,189,34}
  \definecolor{gray127}{RGB}{127,127,127}
  \definecolor{lightgray204}{RGB}{204,204,204}
  \definecolor{mediumpurple148103189}{RGB}{148,103,189}
  \definecolor{orchid227119194}{RGB}{227,119,194}
  \definecolor{sienna1408675}{RGB}{140,86,75}
  \definecolor{steelblue31119180}{RGB}{31,119,180}

  \begin{axis}[
      legend cell align={left},
      legend style={
          fill opacity=0.8,
          draw opacity=1,
          text opacity=1,
          at={(0.03,0.97)},
          anchor=north west,
          draw=lightgray204
        },
      tick align=outside,
      tick pos=left,
      x grid style={darkgray176},
      ylabel={Simulation time ratio $\frac{T_\mathrm{DPOD}}{T_\mathrm{FOM}}$ [-]},
      xlabel={Number of nonlocal damage modes $m_D$ [-]},
      xmin=3.75, xmax=31.25,
      xtick style={color=black},
      y grid style={darkgray176},
      ymin=0.849463083844967, ymax=0.897415896590188,
      ytick style={color=black}
    ]
    \addplot [semithick, mediumpurple148103189, mark=*, mark size=3, mark options={solid}]
    table {%
        5 0.851642757151568
        10 0.853625643356526
        15 0.858394805172531
        20 0.860055971201185
        25 0.864082385210245
        30 0.871871892311915
      };
    \addlegendentry{$m_U = 25$}
    \addplot [semithick, darkturquoise23190207, mark=*, mark size=3, mark options={solid}]
    table {%
        5 0.866948170661597
        10 0.870484046851308
        15 0.87190296609042
        20 0.877820420586023
        25 0.88519823191551
        30 0.895236223283587
      };
    \addlegendentry{$m_U = 50$}
  \end{axis}

\end{tikzpicture}
    \caption{Time ratio of the DPOD simulation compared to the full-order simulation for different numbers of modes $m_U$ and $m_D$.}
    \label{fig: POD split time smiley}
\end{figure}
\subsubsection*{DDEIM}
\begin{figure}[htbp]
    \centering
    \includesvg[width=0.9\textwidth]{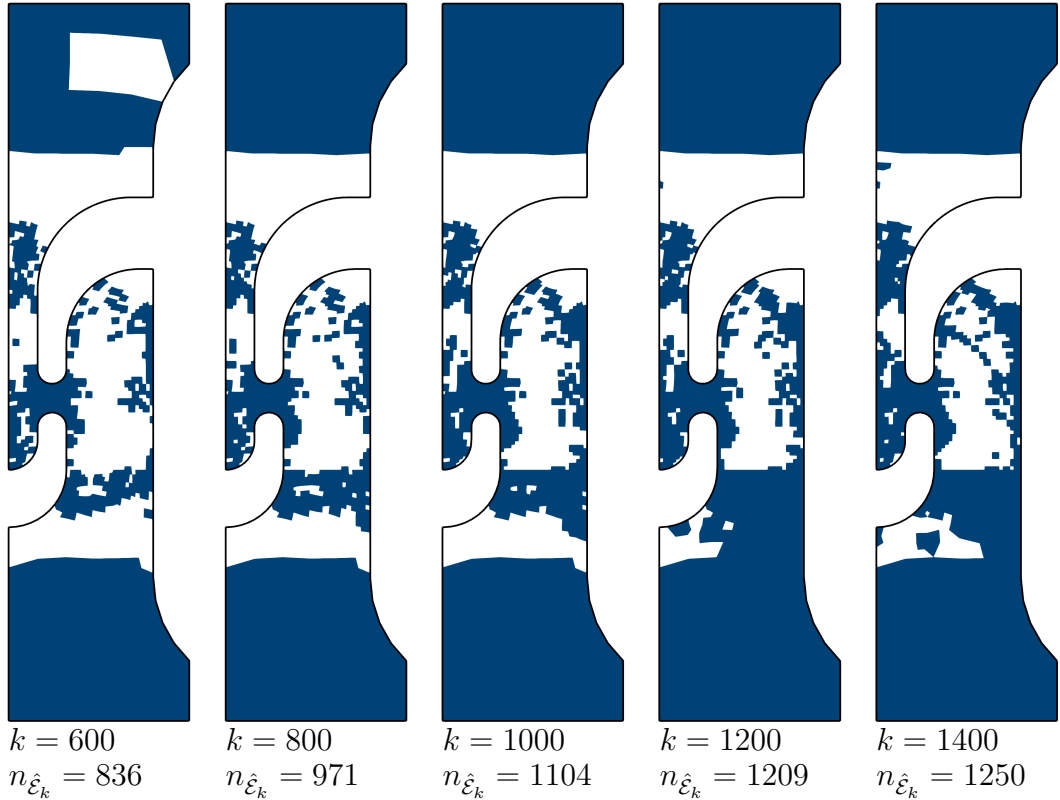}
    \caption{Element selection of the DDEIM for $m_U = 50$, $m_D = 25$ and different number of DDEIM DOFs $k$. The selected elements are shown in blue. The number of selected elements $n_{\hat{\mathcal{E}}_k}$ corresponding to the number of selected DOFs $k$ is indicated in the image.}
    \label{fig: DEIM elem selection smiley}
\end{figure}

The second investigation is concerned with the decomposed DEIM method. The snapshots of the displacements as well as the nonlinear part of the internal forces are obtained from the same simulation as in case of the DPOD method. A study is performed to see the influence of the number of DDEIM DOFs $k$ and, therefore, the number of evaluated elements on the accuracy. The error is computed in the same way as for the DPOD method. The results are shown for $m_U = \{50, 25\}$, $m_D = 25$ in \autoref{fig: DEIM split error smiley}. It can be seen that the error generally decreases with an increasing number of DDEIM DOFs $k$. However, with $k = 600$ DOFs, the simulation did not converge. Only when $k$ is increased to $k = 800$, the error is in the order of $10^{-3}$ using $m_U = 50$ displacement modes and $m_D = 25$ non-local damage modes. With $k = 1000$, the error is in the order of $10^{-4}$, but using a higher number of DDEIM DOFs $k$ does not increase the accuracy anymore. For $m_U=25$ and $m_D = 25$, the error using $k = 1200$ DDEIM DOFs is actually significantly higher than for $k=1000$, meaning it is not guaranteed that more DDEIM DOFs $k$ lead to better results. The simulation time ratio $\frac{T_\mathrm{DDEIM}}{T_\mathrm{FOM}}$ is shown in \autoref{fig: DDEIM split time smiley}. It can be seen that the DDEIM method is able to reduce the simulation time to about 55\% of the full-order simulation time using $k = 1000$ DDEIM DOFs.
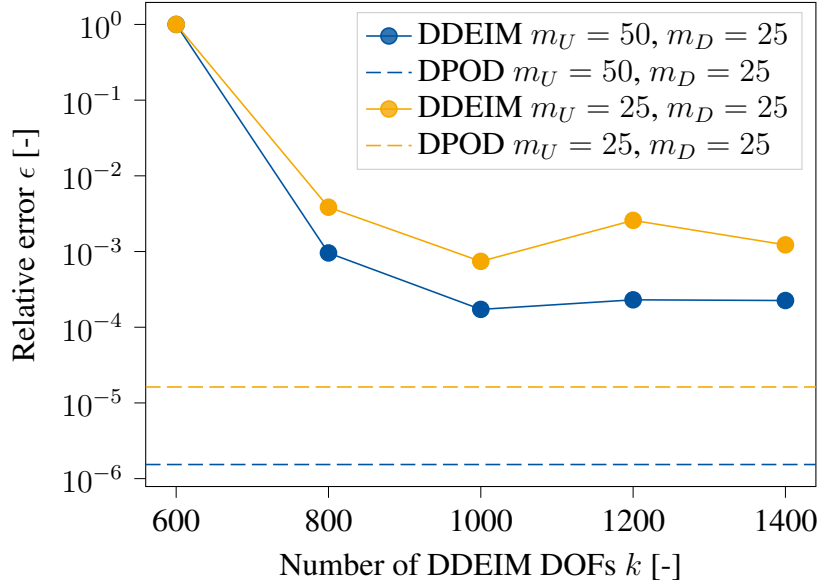
\begin{figure}[htbp]
    \centering
    \pgfplotsset{%
        width=0.65\textwidth,
        height=0.5\textwidth
    }
\begin{tikzpicture}

  \definecolor{darkgray176}{RGB}{176,176,176}
  \definecolor{lightgray204}{RGB}{204,204,204}
  \definecolor{orange2461680}{RGB}{246,168,0}
  \definecolor{teal084159}{RGB}{0,84,159}

  \begin{axis}[
      legend cell align={left},
      legend style={fill opacity=0.8, draw opacity=1, text opacity=1, draw=lightgray204},
      log basis y={10},
      tick align=outside,
      tick pos=left,
      x grid style={darkgray176},
      xlabel={Number of DDEIM DOFs \(\displaystyle k\) [-]},
      xmin=-0.2, xmax=4.2,
      xtick style={color=black},
      xtick={0,1,2,3,4},
      xtick={0,1,2,3,4},
      xtick={0,1,2,3,4},
      xticklabels={600,800,1000,1200,1400},
      y grid style={darkgray176},
      ylabel={Relative error \(\displaystyle \epsilon\) [-]},
      ymin=7.86521453074534e-07, ymax=1.95290199426019,
      ymode=log,
      ytick style={color=black},
      ytick={1e-08,1e-07,1e-06,1e-05,0.0001,0.001,0.01,0.1,1,10,100},
      yticklabels={
          \(\displaystyle {10^{-8}}\),
          \(\displaystyle {10^{-7}}\),
          \(\displaystyle {10^{-6}}\),
          \(\displaystyle {10^{-5}}\),
          \(\displaystyle {10^{-4}}\),
          \(\displaystyle {10^{-3}}\),
          \(\displaystyle {10^{-2}}\),
          \(\displaystyle {10^{-1}}\),
          \(\displaystyle {10^{0}}\),
          \(\displaystyle {10^{1}}\),
          \(\displaystyle {10^{2}}\)
        }
    ]
    \addplot [semithick, teal084159, mark=*, mark size=3, mark options={solid}]
    table {%
        0 1
        1 0.000958063173847695
        2 0.000172085925824881
        3 0.000230186272365567
        4 0.000225373598257047
      };
    \addlegendentry{DDEIM $m_U = 50$, $m_D = 25$}
    \path [draw=teal084159, semithick, dash pattern=on 5.55pt off 2.4pt]
    (axis cs:-0.2,1.53599931423768e-06)
    --(axis cs:4.2,1.53599931423768e-06);
    \addlegendimage{semithick,color=teal084159, dash pattern=on 5.55pt off 2.4pt}
    \addlegendentry{DPOD $m_U = 50$, $m_D = 25$}

    \addplot [semithick, orange2461680, mark=*, mark size=3, mark options={solid}]
    table {%
        0 1
        1 0.00384950343476837
        2 0.000741699269568374
        3 0.00258041293054237
        4 0.00122555416296714
      };
    \addlegendentry{DDEIM $m_U = 25$, $m_D = 25$}
    \path [draw=orange2461680, semithick, dash pattern=on 5.55pt off 2.4pt]
    (axis cs:-0.2,1.62383923765504e-05)
    --(axis cs:4.2,1.62383923765504e-05);
    \addlegendimage{semithick,color=orange2461680, dash pattern=on 5.55pt off 2.4pt}
    \addlegendentry{DPOD $m_U = 25$, $m_D = 25$}

  \end{axis}

\end{tikzpicture}
    \caption{Error of the DDEIM simulation compared to the full-order simulation for different numbers of DDEIM DOFs $k$.}
    \label{fig: DEIM split error smiley}
\end{figure}
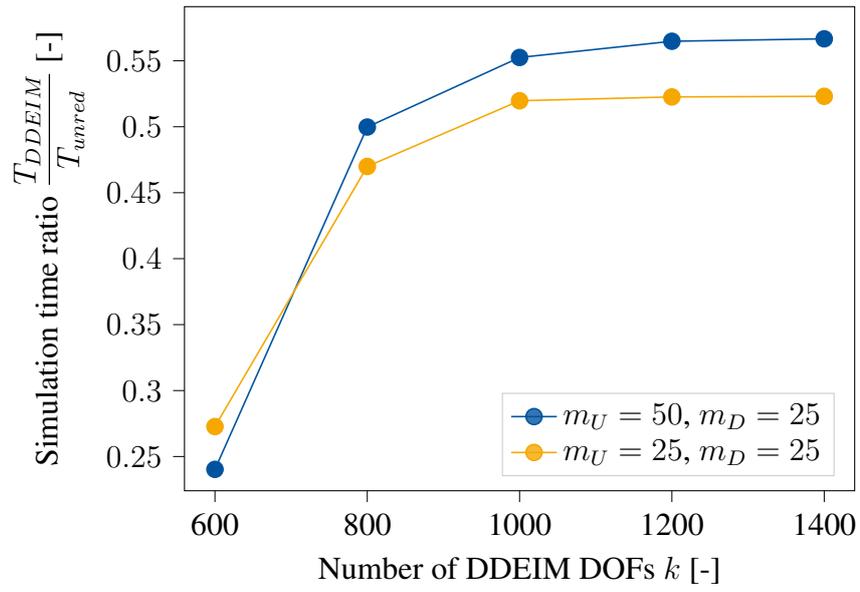
\begin{figure}[htbp]
    \centering
    \pgfplotsset{%
        width=0.65\textwidth,
        height=0.5\textwidth
    }
\begin{tikzpicture}

  \definecolor{darkgray176}{RGB}{176,176,176}
  \definecolor{lightgray204}{RGB}{204,204,204}
  \definecolor{orange2461680}{RGB}{246,168,0}
  \definecolor{teal084159}{RGB}{0,84,159}

  \begin{axis}[
      legend cell align={left},
      legend style={
          fill opacity=0.8,
          draw opacity=1,
          text opacity=1,
          at={(0.97,0.03)},
          anchor=south east,
          draw=lightgray204
        },
      tick align=outside,
      tick pos=left,
      x grid style={darkgray176},
      xlabel={Number of DDEIM DOFs \(\displaystyle k\) [-]},
      xmin=-0.2, xmax=4.2,
      xtick style={color=black},
      xtick={0,1,2,3,4},
      xtick={0,1,2,3,4},
      xtick={0,1,2,3,4},
      xticklabels={600,800,1000,1200,1400},
      y grid style={darkgray176},
      ylabel={Simulation time ratio \(\displaystyle \frac{T_{DDEIM}}{T_{unred}}\) [-]},
      ymin=0.224034169853853, ymax=0.591050494141119,
      ytick style={color=black}
    ]
    \addplot [semithick, teal084159, mark=*, mark size=3, mark options={solid}]
    table {%
        0 0.240262184594184
        1 0.499782714854667
        2 0.552527442523642
        3 0.564822479400789
        4 0.56666900545497
      };
    \addlegendentry{$m_U = 50$, $m_D = 25$}
    \addplot [semithick, orange2461680, mark=*, mark size=3, mark options={solid}]
    table {%
        0 0.272707529508867
        1 0.46991148995635
        2 0.519713393301462
        3 0.52262437665973
        4 0.52307746740593
      };
    \addlegendentry{$m_U = 25$, $m_D = 25$}
  \end{axis}

\end{tikzpicture}
    \caption{Time ratio of the DDEIM simulation compared to the full-order simulation for different numbers of DDEIM DOFs $k$.}
    \label{fig: DDEIM split time smiley}
\end{figure}
The selected elements of the DDEIM for $m_U = 50$, $m_D = 25$ and different numbers of DDEIM DOFs $k$ are shown in \autoref{fig: DEIM elem selection smiley}. It can be seen that the number of selected elements $n_{\hat{\mathcal{E}}_k}$ almost stagnates from $k = 1000$ on. This is due to the fact that DOFs are selected that are already covered by the selected elements.
\subsubsection*{DECSW}
The third investigation is concerned with the decomposed ECSW method. Again, the same snapshots as for the DPOD are used. The error is computed in the same way as for the DPOD method. The results are shown in \autoref{fig: ECSW split error smiley} using $m_U = \{50, 25\}$, $m_D = 25$ and different tolerances $\tau = \{10^{-1}, 10^{-2}, 10^{-3}, 10^{-4}, 10^{-5}, 10^{-6}\}$. It is observed that as the tolerance $\tau$ decreases, the error converges to the error of the DPOD, as could be expected.
\begin{figure}[htbp]
    \centering
    \pgfplotsset{%
        width=0.65\textwidth,
        height=0.5\textwidth
    }
\begin{tikzpicture}

  \definecolor{darkgray176}{RGB}{176,176,176}
  \definecolor{darkorange25512714}{RGB}{255,127,14}
  \definecolor{lightgray204}{RGB}{204,204,204}
  \definecolor{steelblue31119180}{RGB}{31,119,180}
  \definecolor{orange2461680}{RGB}{246,168,0}
  \definecolor{teal084159}{RGB}{0,84,159}

  \begin{axis}[
      legend cell align={left},
      legend style={fill opacity=0.8, draw opacity=1, text opacity=1, draw=lightgray204},
      log basis y={10},
      tick align=outside,
      tick pos=left,
      x grid style={darkgray176},
      xmin=-0.25, xmax=5.25,
      ylabel={Relative error $\epsilon$ [-]},
      xlabel={DECSW tolerance $\tau$ [-]},
      xtick style={color=black},
      xtick={0,1,2,3,4,5},
      xticklabels={$10^{-1}$,$10^{-2}$,$10^{-3}$,$10^{-4}$,$10^{-5}$,$10^{-6}$},
      y grid style={darkgray176},
      ymin=8.48684920288592e-07, ymax=0.0634100283993644,
      ymode=log,
      ytick style={color=black},
      ytick={1e-08,1e-07,1e-06,1e-05,0.0001,0.001,0.01,0.1,1},
      yticklabels={
          \(\displaystyle {10^{-8}}\),
          \(\displaystyle {10^{-7}}\),
          \(\displaystyle {10^{-6}}\),
          \(\displaystyle {10^{-5}}\),
          \(\displaystyle {10^{-4}}\),
          \(\displaystyle {10^{-3}}\),
          \(\displaystyle {10^{-2}}\),
          \(\displaystyle {10^{-1}}\),
          \(\displaystyle {10^{0}}\)
        }
    ]
    \addplot [semithick, teal084159, mark=*, mark size=3, mark options={solid}]
    table {%
        0 0.0380749415113927
        1 0.000485925157013014
        2 4.39089061408457e-05
        3 7.3881886002856e-06
        4 1.41340033001782e-06
        5 1.52226400631532e-06
      };
    \addlegendentry{DECSW $m_U = 50$, $m_D = 25$}
    \path [draw=teal084159, semithick, dash pattern=on 5.55pt off 2.4pt]
    (axis cs:-0.25,1.53599931309212e-06)
    --(axis cs:5.25,1.53599931309212e-06);
    \addlegendimage{semithick,color=teal084159, dash pattern=on 5.55pt off 2.4pt}
    \addlegendentry{DPOD $m_U = 50$, $m_D = 25$}

    \addplot [semithick, orange2461680, mark=*, mark size=3, mark options={solid}]
    table {%
        0 0.0149930425585643
        1 0.000613973095372941
        2 4.24613719425038e-05
        3 1.73081801189993e-05
        4 1.59416325361335e-05
        5 1.62299615307223e-05
      };
    \addlegendentry{DECSW $m_U = 25$, $m_D = 25$}
    \path [draw=orange2461680, semithick, dash pattern=on 5.55pt off 2.4pt]
    (axis cs:-0.25,1.62383923752291e-05)
    --(axis cs:5.25,1.62383923752291e-05);
    \addlegendimage{semithick,color=orange2461680, dash pattern=on 5.55pt off 2.4pt}
    \addlegendentry{DPOD $m_U = 25$, $m_D = 25$}

  \end{axis}

\end{tikzpicture}
    \caption{Error of the DECSW simulation compared to the full-order simulation for different numbers of modes $m_U$ and $m_D$ and different tolerances $\tau$.}
    \label{fig: ECSW split error smiley}
\end{figure}
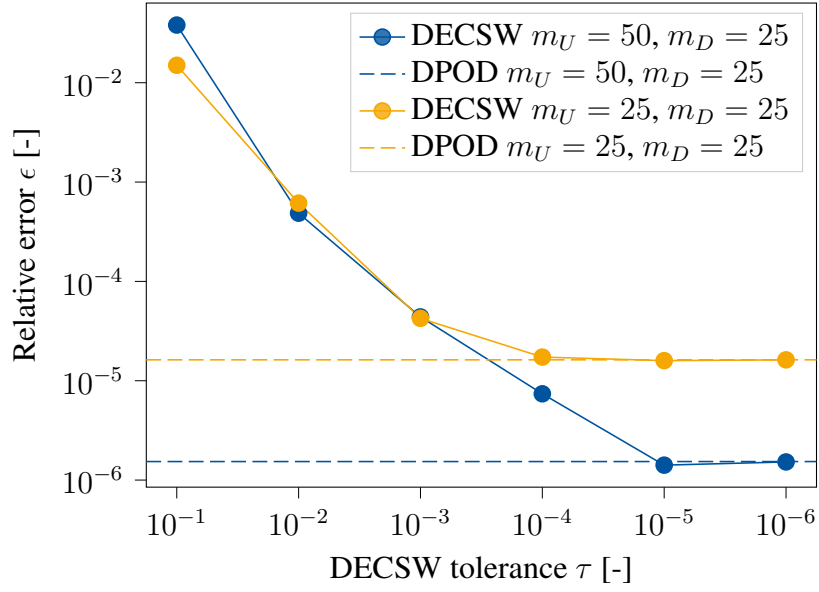
\begin{figure}[htbp]
    \centering
    \pgfplotsset{%
        width=0.65\textwidth,
        height=0.5\textwidth
    }
\begin{tikzpicture}

  \definecolor{darkgray176}{RGB}{176,176,176}
  \definecolor{lightgray204}{RGB}{204,204,204}
  \definecolor{orange2461680}{RGB}{246,168,0}
  \definecolor{teal084159}{RGB}{0,84,159}

  \begin{axis}[
      legend cell align={left},
      legend style={
          fill opacity=0.8,
          draw opacity=1,
          text opacity=1,
          at={(0.03,0.97)},
          anchor=north west,
          draw=lightgray204
        },
      tick align=outside,
      tick pos=left,
      x grid style={darkgray176},
      xmin=-0.25, xmax=5.25,
      ylabel={Simulation time ratio $\frac{T_\mathrm{DECSW}}{T_\mathrm{FOM}}$ [-]},
      xlabel={DECSW tolerance $\tau$ [-]},
      xtick style={color=black},
      xtick={0,1,2,3,4,5},
      xticklabels={$10^{-1}$,$10^{-2}$,$10^{-3}$,$10^{-4}$,$10^{-5}$,$10^{-6}$},
      y grid style={darkgray176},
      ymin=0.0611867688747975, ymax=0.622687822052824,
      ytick style={color=black}
    ]
    \addplot [semithick, teal084159, mark=*, mark size=3, mark options={solid}]
    table {%
        0 0.114894962406377
        1 0.159167793642103
        2 0.223090972081414
        3 0.317660181027965
        4 0.461707782139659
        5 0.597165046908368
      };
    \addlegendentry{$m_U = 50$, $m_D = 25$}
    \addplot [semithick, orange2461680, mark=*, mark size=3, mark options={solid}]
    table {%
        0 0.0867095440192533
        1 0.124322199914059
        2 0.173130855983883
        3 0.257813353266655
        4 0.422164494530287
        5 0.564666601976677
      };
    \addlegendentry{$m_U = 25$, $m_D = 25$}
  \end{axis}

\end{tikzpicture}
    \caption{Time ratio of the DECSW simulation compared to the full-order simulation for different numbers of modes $m_U$ and $m_D$ and different tolerances $\tau$.}
    \label{fig: ECSW split time smiley}
\end{figure}
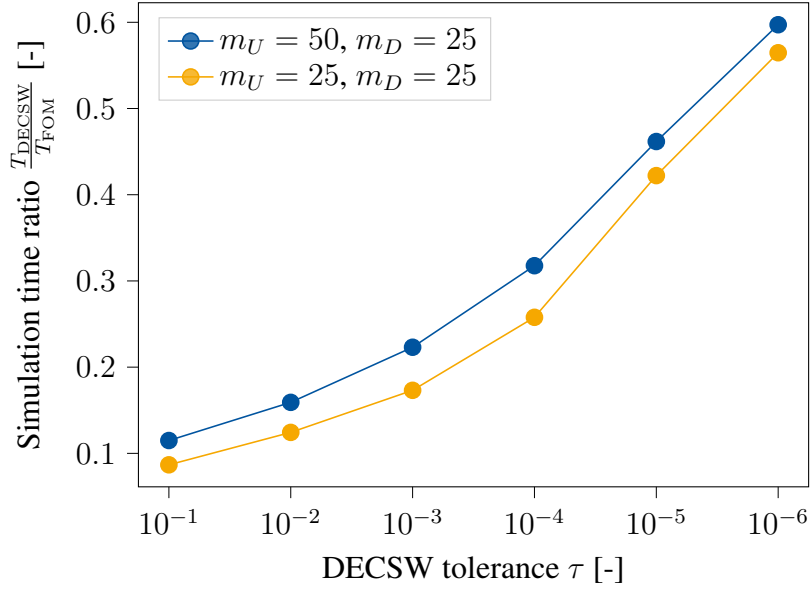
\begin{figure}[htbp]
    \centering
    \includesvg[width=\textwidth]{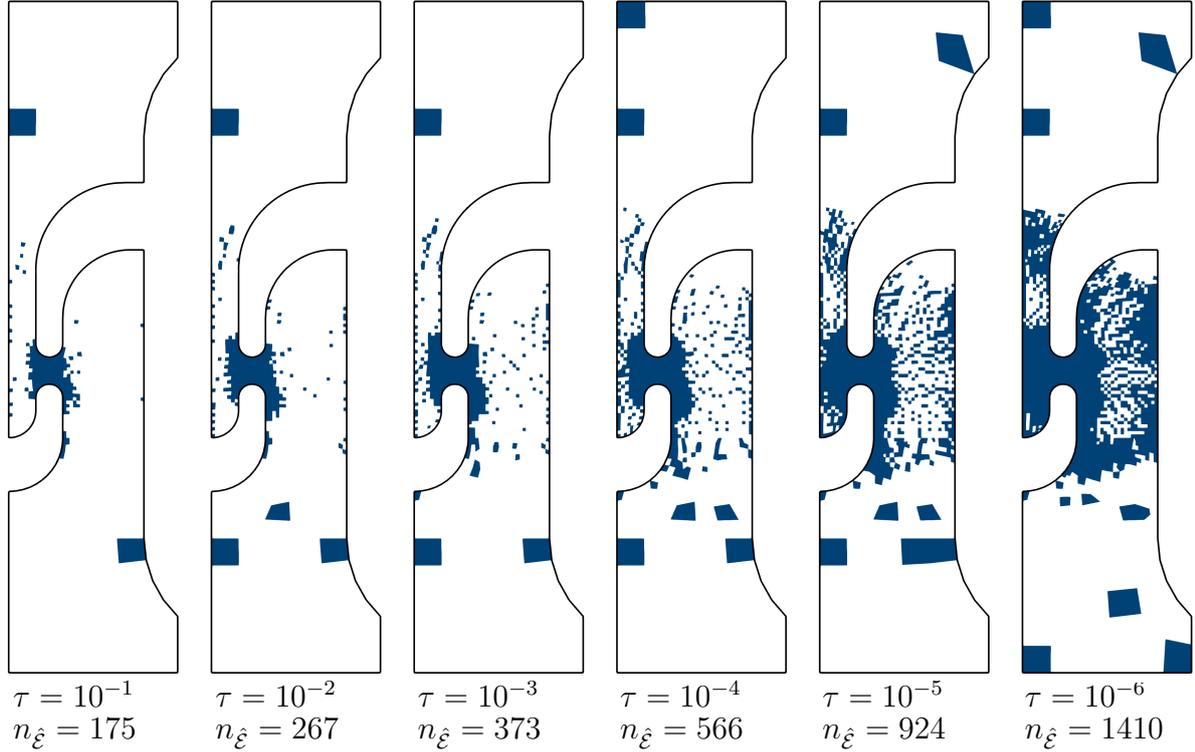}
    \caption{Element selection of the ECSW for $m_U = 50$, $m_D = 25$ and different tolerances $\tau$. The selected elements are shown in blue. The number of selected elements $n_{\hat{\mathcal{E}}}$ corresponding to the tolerance $\tau$ is indicated in the image.}
    \label{fig: ECSW elem selection smiley}
\end{figure}
The simulation time ratio $\frac{T_\mathrm{DECSW}}{T_\mathrm{FOM}}$ is shown in \autoref{fig: ECSW split time smiley}. It can be seen that the DECSW method is able to reduce the simulation time to about 15\% of the full-order simulation time if the tolerance $\tau$ is set to $10^{-2}$, where the error is still in the order of $10^{-3}$. If higher accuracy is required, the simulation time can be reduced to about 47\% of the full-order simulation time with an error in the order of $10^{-6}$ using a tolerance of $\tau = 10^{-5}$. The selected elements and the number of selected elements $n_{\hat{\mathcal{E}}}$ for the ECSW method are shown in \autoref{fig: ECSW elem selection smiley} for $m_U = 50$, $m_D = 25$ and different tolerances $\tau$. Interestingly, the element selection of the DECSW looks very different than the element selection of the DDEIM. Whereas the DDEIM algorithm selects many elements near the lower and upper edge of the structure, the DECSW algorithm chooses significantly more elements in the center of the structure, where damage is evolving. Already for a tolerance of $\tau = 10^{-1}$, the damaged zone is nearly fully covered by the selected elements.

\subsubsection*{Prediction}
\begin{figure}[htbp]
    \centering
    \pgfplotsset{%
        width=0.65\textwidth,
        height=0.5\textwidth
    }
    \input{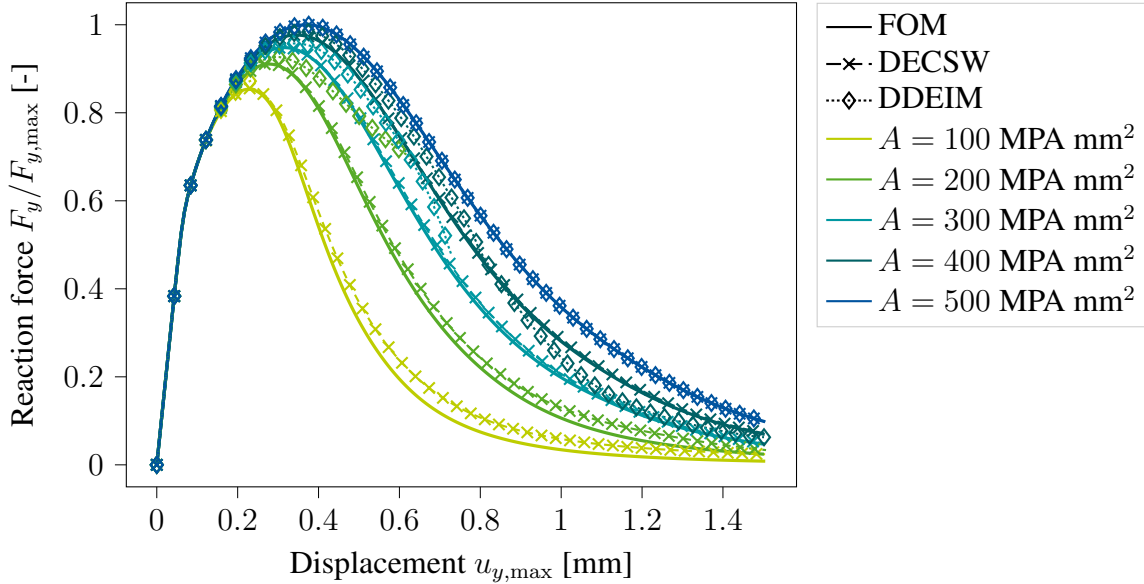}
    \caption{Comparison between the predictions using the DECSW method with $m_U = 50$ displacement modes, $m_D = 25$ non-local damage modes, and $n_{\hat{\mathcal{E}}} = 924$ out of $2649$ elements, using the DDEIM with $m_U = 50$ displacement modes, $m_D = 25$ non-local damage modes, and $n_{\hat{\mathcal{E}}} = 1104$ out of $2649$ elements, and the full-order reference simulation for different values of the gradient parameter $A$. The reduced order models are created using only the snapshots from the full-order simulation with $A = 500$ MPa mm$^{\mathrm{2}}$. Markers are only shown for every 5th point for better visibility.}
    \label{fig: ECSW split A100}
\end{figure}

\begin{figure}
    \centering
    \includesvg[width=\textwidth]{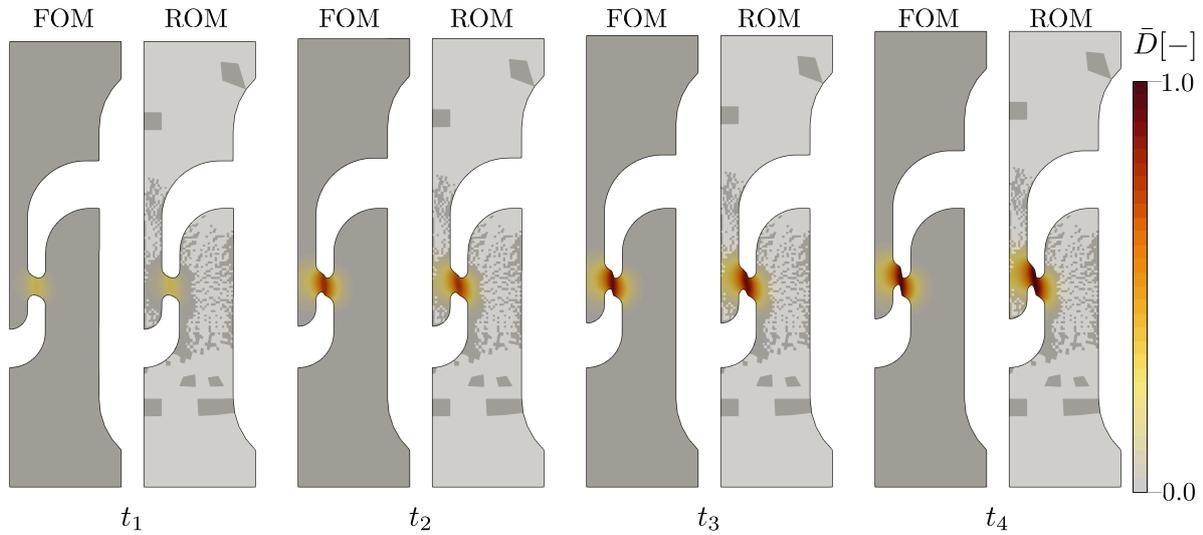}
    \caption{Comparison of the non-local damage variable at four pseudo-timesteps $t_1$, $t_2$, $t_3$, and $t_4$ for the full-order simulation and the DECSW with $m_U = 50$, $m_D = 25$ and $\tau = 10^{-5}$.}
    \label{fig: Dbar A100}
\end{figure}

Up to this point, only a reconstruction of the full-order solution used for creating the reduced order models has been investigated. Now, to investigate the prediction capabilities of the methods, the same snapshots as before are used, but parameter $A$ controlling the influence of the gradient term in the formulation is reduced, which leads to a much narrower damage zone. The resulting force-displacement curves using $m_U = 50$, $m_D = 25$ and $\tau = 10^{-5}$ for the DECSW and $m_U = 50$, $m_D = 25$ and $k = 500$ for the DDEIM are shown in \autoref{fig: ECSW split A100} for $A = \{100, 200, 300, 400, 500\}$ MPa mm$^{2}$. The force-displacement curves of the full-order simulations are shown for reference. For $A=400$, the DDEIM completes the simulation, but the difference to the full-order reference simulation is significant. For $A = 300$ MPa mm$^{2}$ and lower, the DDEIM suffers from instabilities that lead to a termination of the simulations. For these cases, the limit (i.e. maximum) load is also not approximated well.
Qualitatively, the DECSW method is able to capture the force-displacement curves well and approximate the limit load with high accuracy. Nevertheless, small differences in the softening regime are visible. A comparison of the non-local damage variable between the DECSW and the full-order reference simulation at four different pseudo-timesteps is shown in \autoref{fig: Dbar A100} for $A = 100$ MPA mm$^{2}$. The narrower damage pattern is captured well, but due to the more distributed damage in the snapshots, the damage variable is slightly overestimated in the vicinity of the primary damage zone.

\subsection{Three-dimensional plate with hole}
\begin{figure}
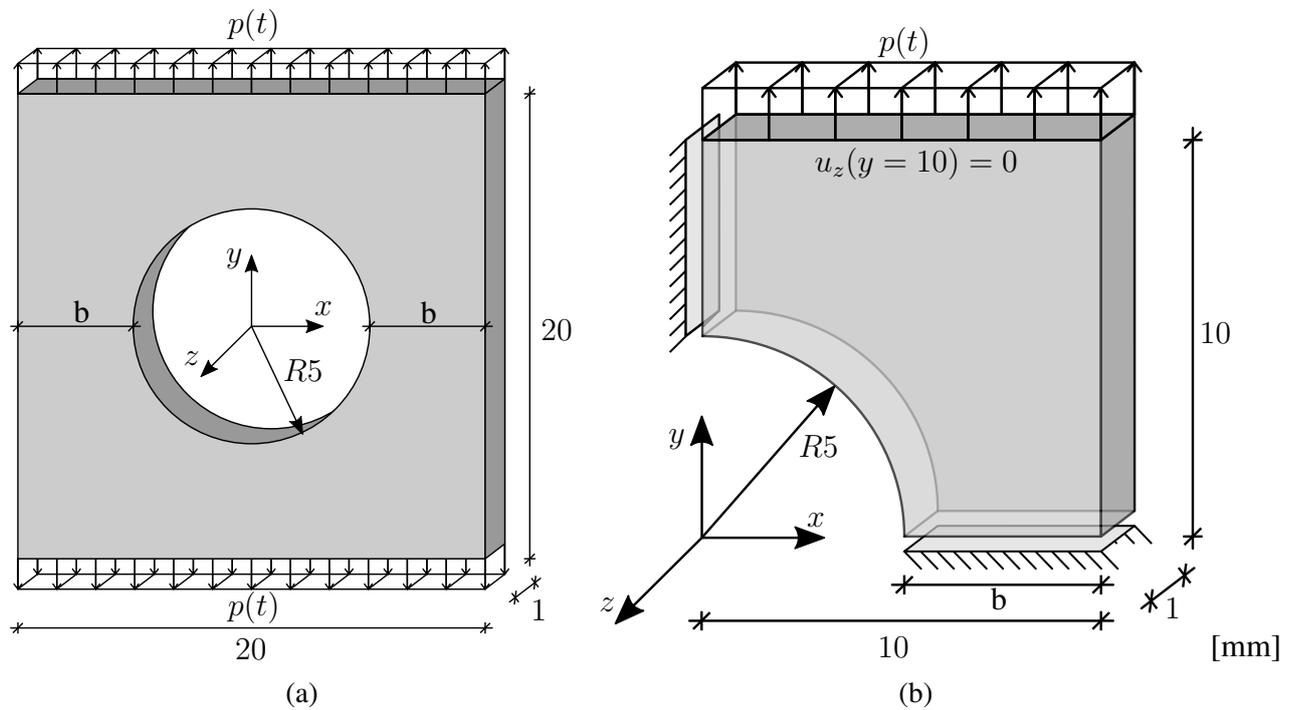

    \begin{subfigure}[t]{0.48\textwidth}
        \centering
        \includesvg[width=\textwidth]{images/plate.svg}
        \caption{}
        \label{fig: plate with hole}
    \end{subfigure}
    \begin{subfigure}[t]{0.52\textwidth}
        \centering
        \includesvg[width=\textwidth]{images/platequarter.svg}
        \caption{}
        \label{fig: plate with hole sym}
    \end{subfigure}
    \caption{(a) Original geometry of the plate with a hole. (b) Final boundary value problem for the plate with a hole (symmetry exploited).}
    \label{fig: plate hole}
\end{figure}

\begin{figure}
    \centering
    \includesvg[width=\textwidth]{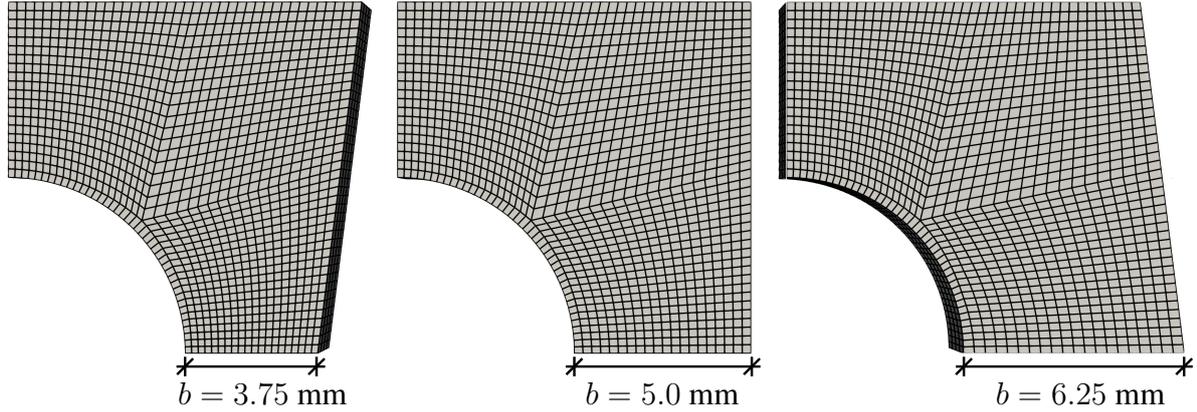}
    \caption{Discretization into $4804$ trilinear hexahedral finite elements for three different widths $b=\{3.75, 5.0, 6.25\}$ mm.}
    \label{fig: plate hole discretization}
\end{figure}

\begin{figure}[htbp]
    \centering
    \pgfplotsset{%
        width=0.65\textwidth,
        height=0.5\textwidth
    }
    \input{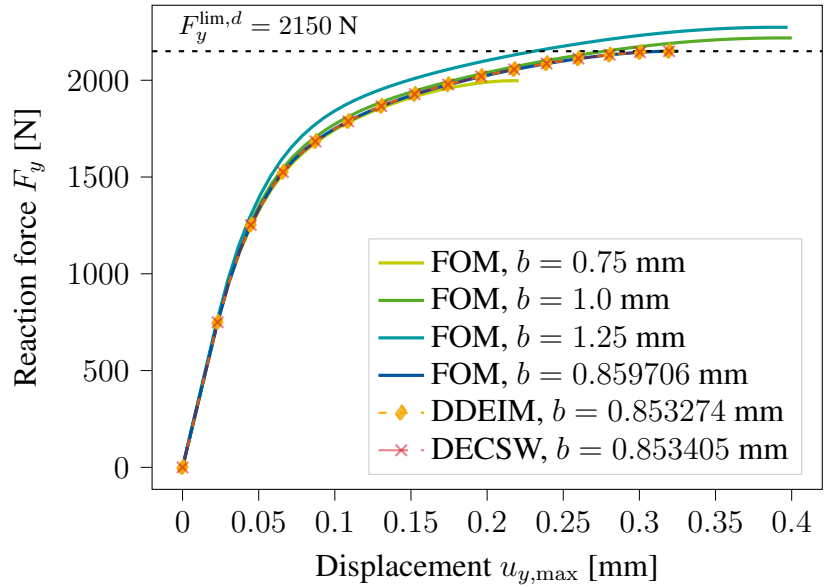}
    \caption{Comparison of the force-displacement curves for the optimal widths $b$ obtained from the DDEIM and DECSW methods with the full-order simulation. The optimal width $b$ is computed for a desired limit load of $F_y^{\text{lim}, d} = 2150$ N.}
    \label{fig: plate hole optimization force disp}
\end{figure}

To test the methods' applicability to an optimization problem, a three-dimensional plate with circular hole is considered, where the width of the plate $b$ at $y=0$ mm is varied. The boundary value problem is shown in \autoref{fig: plate with hole}.
The discretization into $4804$ trilinear hexahedral finite elements is shown exemplarily for three different widths $b=\{3.75, 5.0, 6.25\}$ mm in \autoref{fig: plate hole discretization}. The discretization was shown on the basis of a mesh convergence study.
The simulation is performed until the limit load is reached. The optimization problem is to find the width $b$ for a desired limit load $F_y^{\text{lim}, d}$. The optimization problem is solved using \textit{Brent's method} \citep{brent_efficient_1973}, which is a well-established root-finding algorithm and built into the \textit{Python} package \textit{SciPy} \citep{virtanen_scipy_2020}. The widths $b = 3.75, 5.0, 6.25$ mm are used as initial guesses for the optimization problem. The desired limit load is set to $F_y^{\text{lim}, d} = 2150$ N. The optimization problem is solved using both the DDEIM and the DECSW method. The three initial steps are computed with the full-order model to obtain the snapshots for the reduced-order methods. The results of the optimization problem are shown in \autoref{tab: plate hole optimization}.
\begin{table}[htbp]
    \centering
    \caption{Results of the optimization problem for the width $b_i$ [mm] of the plate with a hole. The optimization is performed using the DDEIM and DECSW methods introduced in Sections \ref{sec: DDEIM} and \ref{sec: ECSWM}, respectively. The first three iteration steps are always computed using the FOM and the snapshots are used to compute the ROMs. The optimization error is computed as $\epsilon = (F_y^{\text{lim}} - F_y^{\text{lim}, d})^2$.}
    \label{tab: plate hole optimization}
    \renewcommand{\arraystretch}{1.2}
    \begin{tabular}{r|ll|ll|ll}
                      & \multicolumn{2}{c|}{\begin{tabular}[c]{@{}c@{}}FOM\\ $n_{\mathcal{E}} = 4804$\end{tabular}} & \multicolumn{2}{c|}{\begin{tabular}[c]{@{}c@{}}DDEIM\\ $m_U, m_D = 50, k = 400$\\ $n_{\hat{\mathcal{E}}_k} = 1598$\end{tabular}} & \multicolumn{2}{c}{\begin{tabular}[c]{@{}c@{}}DECSW\\ $m_U, m_D = 50, \tau = 10^{-3}$\\ $n_{\hat{\mathcal{E}}} = 801$\end{tabular}}                                                         \\ \hline
        Iteration $i$ & Width $b_i$                                     & Error $\epsilon_i$                              & Width $b_i$                                    & Error $\epsilon_i$ & Width $b_i$ & Error $\epsilon_i$ \\ \hline
        1             & 0.75                                            & 23181.23                                        & 0.75                                           & 23181.23           & 0.75        & 23181.23           \\
        2             & 1.0                                             & 4702.298                                        & 1.0                                            & 4702.298           & 1.0         & 4702.298           \\
        3             & 1.25                                            & 15308.95                                        & 1.25                                           & 15308.95           & 1.25        & 15308.95           \\ \hline
        4             & 0.904508                                        & 1126.917                                        & 0.904508                                       & 1551.595           & 0.904508    & 1456.776           \\
        5             & 0.845492                                        & 190.5739                                        & 0.845492                                       & 64.46864           & 0.845492    & 61.30820           \\
        6             & 0.818193                                        & 2044.616                                        & 0.809017                                       & 3152.794           & 0.809017    & 3042.660           \\
        7             & 0.866827                                        & 40.22574                                        & 0.864154                                       & 102.2571           & 0.864287    & 94.96781           \\
        8             & 0.861954                                        & 4.006129                                        & 0.853036                                       & 0.013807           & 0.853241    & 0.012558           \\
        9             & 0.860165                                        & 0.164554                                        & 0.853758                                       & 0.201927           & 0.853870    & 0.198040           \\
        10            & 0.859706                                        & 0.000071                                        & 0.853274                                       & 0.001212           & 0.853405    & 0.001108           \\
        11            & 0.859792                                        & 0.006879                                        & 0.853360                                       & 0.012019           & 0.853490    & 0.011895           \\
        12            & 0.859620                                        & 0.004373                                        & 0.853189                                       & 0.013403           & --          & --                 \\ \hline
        Time {[}s{]}  & \multicolumn{2}{c|}{7355}                       & \multicolumn{2}{c|}{3503}                       & \multicolumn{2}{c}{2550}
    \end{tabular}
\end{table}
It can be seen that both methods are able to solve the optimization problem much faster than using only the full-order model. It should be noted that the time to create the respective reduced-order models is included in the total time. The DDEIM is able to solve the optimization problem in about 47\% of the time of the full-order simulation, while the DECSW is able to solve the optimization problem in about 35\% of the time of the full-order simulation. From the found widths $b$, the relative approximation error of the ROMs is calculated. The optimal width $b$ computed by the DDEIM in iteration $10$ has a relative error of $0.75\%$, whereas the width computed using the DECSW method in iteration $10$ has a relative error of $0.73\%$. The resulting force-displacement curves for the optimal widths $b$ are shown in \autoref{fig: plate hole optimization force disp}.

\section{Conclusion}
\label{sec: conclusion}

In this work, two distinct decomposed hyper-reduced order modeling approaches for general continuum mechanical multi-field problems were presented for the first time, illustrated at the example of a coupled gradient-extended damage-plasticity model. Building on the previously introduced multi-field decomposition for POD, called DPOD (see \citet{zhang_multi-field_2025}), both the DEIM and the ECSW method were extended to account for the multi-field nature of the problem, leading to so-called DDEIM and DECSW methods, respectively. Using the presented approaches, the formerly observed instabilities of the original non-decomposed methods in the context of highly nonlinear damage-plasticity simulations involving severe softening could be circumvented. Two numerical examples were used to demonstrate the performance of the new methods. In the first investigation of the so-called \textit{smiley specimen}, it was shown that although the DPOD is able to capture the nonlinear behavior of the problem with high accuracy, the speedup of the simulation was severely limited by the time needed to compute the tangential stiffness matrix and the residual vector. Both the DDEIM and the DECSW method were able to significantly reduce the simulation time while maintaining a high accuracy. The DECSW showed overall a better performance in terms of accuracy and speedup compared to the DDEIM. In predictive simulations, the DDEIM led to unstable simulations if the solution was too far away from the snapshot data. To the contrary, the DECSW method was stable in this case and able to provide an accurate approximation of the limit load, although some quantitative discrepancies in the softening regime were visible.

The second numerical example focused on a three-dimensional plate with a hole under tension. The width of the plate was parametrized and an optimization was performed to find the optimal width of the plate for a desired limit load. Both methods were able to provide a good approximation of the optimal width. The DECSW method was able to reduce the time of the optimization, including full-order simulations and the computations using the reduced order model, to $35\%$ of the time needed for the conventional full-order optimization. The DDEIM was able to reduce the time to $47\%$ of the time needed for the full-order optimization.

In future work, the DDEIM will be extended to incorporate an unassembled version of the discrete empirical interpolation, which promises to further reduce the simulation time and accuracy. Furthermore, it needs to be investigated, whether the presented multi-field decomposed hyper-reduced order modeling approaches can be applied to problems with more than two fields, such as thermo-mechanical damage-plasticity problems.

\section*{Acknowledgements}
J. Kehls, Q. Zhang, S. Reese, and T. Brepols thankfully acknowledge the funding of project CRC/TRR 339 (subproject B05, project number 453596084) by the DFG. S. Ritzert and T. Brepols acknowledge the funding of project CRC/TRR 280 (subproject A01, project number 417002380) by the DFG.
Furthermore, T. Brepols acknowledges the funding of the project "Accelerating Microstructure-to-Property Computations of Metallic Materials through Physics-Enhanced Deep Learning Techniques - with Special Application to Nickel-Based Superalloys" (project number 561202254) by the DFG.



\bibliographystyle{plainnatshort}
\bibliography{Paper01}

\end{document}